\documentclass[12pt]{article}
\pdfoutput=1
\usepackage{jheppub}
\usepackage{slashed}
\usepackage{amsmath}
\usepackage{amsfonts}
\usepackage{amssymb}
\usepackage{graphicx}

\usepackage[export]{adjustbox}
\newcommand{\bmat}{\left(\begin{array}}
\newcommand{\emat}{\end{array}\right)}

\def\yzero{\smash{\hbox{$y\kern-4pt\raise1pt\hbox{${}^\circ$}$}}}

\def\beq{\begin{equation}}
\def\eeq{\end{equation}}
\def\beqa{\begin{eqnarray}}
\def\eeqa{\end{eqnarray}}

\def\-{\hphantom{-}}
\def\ov{\overline}
\def\s2{\frac{1}{\sqrt2}}

\def\beq{\begin{equation}}
\def\eeq{\end{equation}}
\def\beqa{\begin{eqnarray}}
\def\eeqa{\end{eqnarray}}
\def\tr{{\rm tr \,}}

\def\II{\relax{\rm I\kern-.18em I}}

\def\Dsl{\,\raise.15ex\hbox{/}\mkern-13.5mu D} 
\def\aD9{{\ov{\rm D9}}}

\def\IS{{\bf {S}}}
\def\IR{{\bf {R}}}
\def\IZ{{\bf {Z}}}

\def\IT{{\bf {T}}}

\catcode`\@=11   
\newdimen\@rotdimen
\newbox\@rotbox  

\def\@vspec#1{\special{ps:#1}}
\def\@rotstart#1{\@vspec{gsave currentpoint currentpoint translate
   #1 neg exch neg exch translate}}
\def\@rotfinish{\@vspec{currentpoint grestore moveto}}
\def\@rotr#1{\@rotdimen=\ht#1\advance\@rotdimen by\dp#1%
   \hbox to\@rotdimen{\hskip\ht#1\vbox to\wd#1{\@rotstart{90 rotate}%
   \box#1\vss}\hss}\@rotfinish}
\def\@rotl#1{\@rotdimen=\ht#1\advance\@rotdimen by\dp#1%
   \hbox to\@rotdimen{\vbox to\wd#1{\vskip\wd#1\@rotstart{270 rotate}%
   \box#1\vss}\hss}\@rotfinish}%
\def\@rotu#1{\@rotdimen=\ht#1\advance\@rotdimen by\dp#1%
   \hbox to\wd#1{\hskip\wd#1\vbox to\@rotdimen{\vskip\@rotdimen
   \@rotstart{-1 dup scale}\box#1\vss}\hss}\@rotfinish}%
\def\@rotf#1{\hbox to\wd#1{\hskip\wd#1\@rotstart{-1 1 scale}%
   \box#1\hss}\@rotfinish}%
\def\rotate{\@ifnextchar[{\@rotate}{\@rotate[l]}}
\def\@rotate[#1]#2{\setbox\@rotbox=\hbox{#2}\@nameuse{@rot#1}\@rotbox}

\catcode`\@=12

\begin{document}

\makeatletter
\@addtoreset{equation}{section}
\makeatother
\renewcommand{\theequation}{\thesection.\arabic{equation}}
\pagestyle{empty}
\rightline{IFT-UAM/CSIC-26-076 }
\rightline{ZMP-HH/26-16}
\vspace{.5cm}
\begin{center}
\Large{\bf The Art of Networking:}\\
\Large {\bf Networks of Trivalent 10d Heterotic Junctions}
\\[8mm] 

\large{Chiara Altavista${}^1$, Edoardo Anastasi${}^1$, \\Roberta Angius${}^2$, Angel M. Uranga${}^1$, Chuying Wang${}^1$ \\[4mm]}
\footnotesize{${}^1$ Instituto de F\'{\i}sica Te\'orica IFT-UAM/CSIC,\\[-0.3em] 
C/ Nicol\'as Cabrera 13-15, 
Campus de Cantoblanco, 28049 Madrid, Spain}\\ 
\footnotesize{${}^2$ II. Institut f\"ur Theoretische Physik, Universit\"at Hamburg, \\ Notkestrasse 9, 22607 Hamburg, Germany}\\ 
\footnotesize{\href{mailto:chiara.altavista@estudiante.uam.es}{chiara.altavista@estudiante.uam.es}}, \href{mailto:edo.anastasi@virgilio.it}{edo.anastasi@virgilio.it}, \\\href{mailto:roberta.angius@uni-hamburg.de}{roberta.angius@uni-hamburg.de}, \href{mailto:angel.uranga@csic.es}{angel.uranga@csic.es}, \href{chuying.wang@ift.csic.es}{chuying.wang@ift.csic.es}

\vspace*{8mm}

\small{\bf Abstract} \\
\end{center}
\begin{center}
\begin{minipage}[h]{\textwidth}
\small{We initiate the study of networks of 10d string theories connected by junctions implied by the cobordism conjecture. Focusing on the recently constructed junction of the three 10d non-tachyonic heterotic theories, we generalize its $(0,1)$ heterotic worldsheet description to construct arbitrary networks. For one-dimensional networks, we formulate their topology in terms of graph theory and provide a simple worldsheet realization for general graphs. We then extend our analysis to higher-dimensional networks, describing e.g. nucleation in a theory of bubbles of pairs of other theories. We also discuss compact configurations, which define a novel class of compactifications in which different sectors propagate on different compact spaces, in a way reminiscent of compactifications on quantum geometries like $\IS^1\vee\IS^1$. }

\newpage

\end{minipage}
\end{center}
\newpage
\setcounter{page}{1}
\pagestyle{plain}
\renewcommand{\thefootnote}{\arabic{footnote}}
\setcounter{footnote}{0}

\vspace*{-1cm}

\tableofcontents

\vspace*{1cm}


\section{Introduction}
\label{sec:intro}

The cobordism conjecture \cite{McNamara:2019rup} is one of the cornerstones of recent progress in string theory (both within the swampland program \cite{Vafa:2005ui} and more broadly). It states the trivality of all (suitably defined) cobordism charges of any consistent quantum gravity configuration, and predicts a rich set of spacetime dynamical configurations connecting different theories. Focusing on the most interesting case of uncompactified theories, it implies the existence of
end of the world (ETW) boundaries at which spacetime terminates, like the Ho\v{r}ava-Witten wall in 11d M-theory \cite{Horava:1995qa,Horava:1996ma} (see also \cite{Montero:2025ayi}), or the type I' O8-planes (plus D8-branes) \cite{Polchinski:1995df} for (possibly massive) 10d type IIA theory\footnote{For other dynamical realizations of cobordism ETW boundaries in diverse contexts, see also \cite{Buratti:2021yia,Buratti:2021fiv,Angius:2022aeq,Blumenhagen:2022mqw,Angius:2022mgh,Angius:2023xtu,Blumenhagen:2023abk,Angius:2023uqk,Angius:2024zjv,Delgado:2023uqk,Calderon-Infante:2026ymy,Makridou:2026jzy}.}. It also implies 9d interfaces between 10d string theories, like the recently studied 10d type IIA/IIB domain wall (and its 10d 0A/0B cousin) \cite{Heckman:2025wqd,Anastasi:2026cus,Torres:2026vxx}. 

Recently, further configurations implied by the cobordism conjecture have been constructed in \cite{Altavista:2026edv} and \cite{Tachikawa:2026top} (see also \cite{Altavista:2026brr}), consisting of junctions of multiple 10d string theories. They are basically constructed from a worldsheet perspective, by a generalization of the realization of spacetime cobordisms as processes of `going up and down the RG flow' in the 2d worldsheet theory \cite{Gaiotto:2019asa}.

It is a natural question to consider whether these junctions can be combined into more involved networks of 10d string theories. Conversely, whether high-valence junctions can be resolved into a network of several lower-valence ones, and hence whether there is a unique junction (or a finite set of them) which is the fundamental building block of any network. 

One might think that the answers to these questions are obvious, or even tautological, since any junction configuration is topologically allowed because it relates theories in the trivial cobordism class. However, there is a precise context in which these questions admit highly non-trivial answers. Even if all junctions are topologically allowed by the cobordism conjecture, they generically require strong coupling, in particular when they involve chiral theories whose chiral fields must be gapped at the junction. In contrast, \cite{Altavista:2026edv} uncovered the existence (and built several examples) of junctions (dubbed bouquets) in which the chiral fields in the theory in one branch flow across the junction into the other branches, a pattern which we refer to as chiral flow. For this kind of configurations, no strong coupling is required, and furthermore the junction is topologically protected against splitting (in contrast with junctions with trivial chiral flow, which can split into disjoint components). Hence, a well defined and non-trivial problem is that of the structure of networks based on bouquets of 10d chiral string theories. 

\begin{figure}[htb]
\begin{center}
\includegraphics[scale=.35]{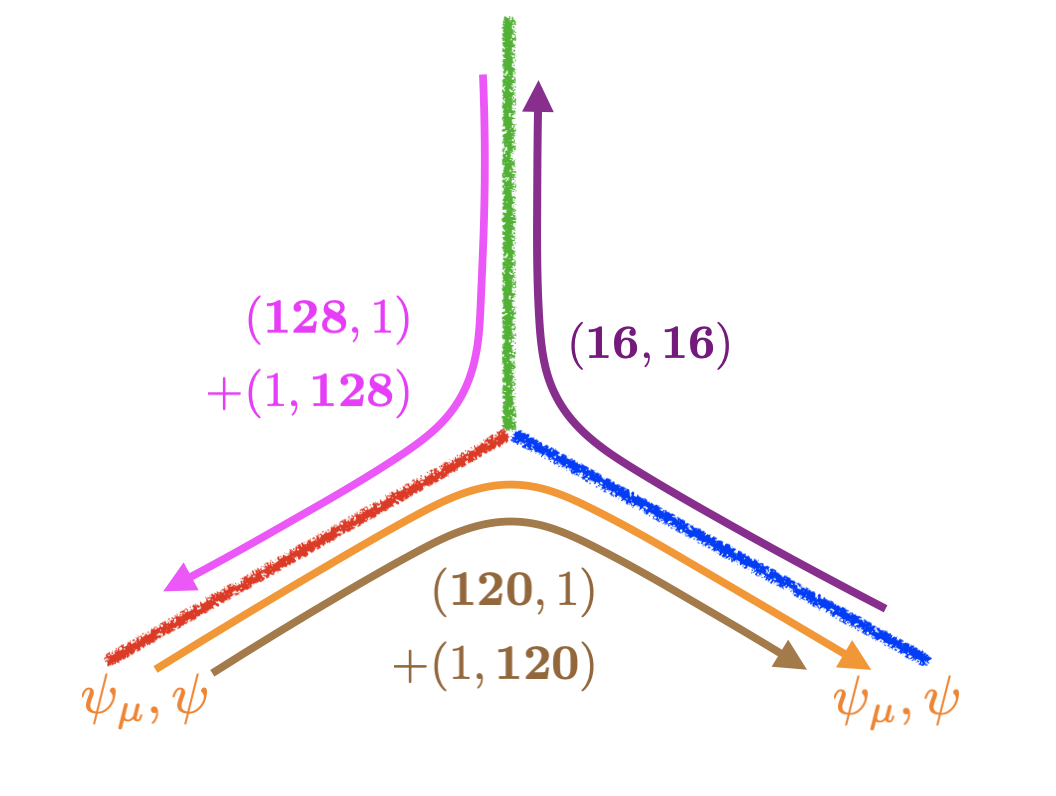}
\caption{\small Spacetime picture of the bouquet of the three non-tachyonic heterotic strings, with the flow of chiral fields among branches (Adapted from \cite{Altavista:2026edv}).}
\label{fig:bouquet-chiral-flow}
\end{center}
\end{figure}

In this work we initiate the analysis of such networks, focusing on a case study.  We focus on networks built from a remarkable bouquet of the three 10d non-tachyonic heterotic theories, proposed in \cite{Altavista:2026edv}, and constructed explicitly shortly after in \cite{Tachikawa:2026top} (see also \cite{Altavista:2026brr}). The junction displays a beautiful chiral flow, shown in Figure \ref{fig:bouquet-chiral-flow}. The construction in \cite{Tachikawa:2026top} used a remarkably simple $(0,1)$ heterotic worldsheet theory, in terms of two sets of additional fields (to move up and down the RG flow), a 2d $\IZ_2$ gauge symmetry, and a superpotential constraining the dynamics to a locus with the topology of a junction.

We will show that a generalization of this construction allows to glue junctions together to build networks of 10d theories. For one-dimensional networks, we will provide the topological building rules in terms of graph theory, and provide a simple characterization of the topology of general networks. Finally, we show that graph theoretical concepts allow to translate this characterization into a simple worldsheet  description of a general graph. We also discuss higher-dimensional networks, and their possible physical interpretation as creation of bubbles of other theories. Finally, we argue that compact networks can be regarded as novel compactifications, in which different fields propagate on different compact spaces. This bears a strong resemblance with the recent proposal of compactifications of fields with different boundary conditions on the quantum geometry $\IS^1\vee\IS^1$ \cite{Baykara:2026gem,Altavista:2026evd,Baykara:2026vdc,Altavista:2026brr,Dasgupta:2026maq,Basile:2026trt,Kamal:2026msr}, although the precise relation remains an open question.

An important aspect of our discussion is that, unlike the basic junction, the more general networks are not topologically protected against decay into simpler configurations. At first sight, this may appear to diminish the value of a systematic characterization of the full space of networks and their worldsheet realizations. Our perspective, however, is that establishing such a characterization constitutes a necessary first step toward understanding whether and how these systems may be stabilized, for instance through the introduction of additional branes, fluxes, or other ingredients.
A useful analogy is provided by the history of other non-supersymmetric constructions, such as brane–antibrane systems. Despite their apparent instability, their detailed study ultimately transformed antibranes into a standard ingredient of string-theoretic model building and led to important advances in our understanding of D-branes and RR fields in terms of K-theory \cite{Witten:1998cd}. We believe that networks of string theories may likewise reveal novel structures and mechanisms whose significance is not yet apparent. While the present work should be viewed as an exploratory step, we regard the potential insights offered by a systematic study of these configurations as sufficient motivation to investigate them despite their lack of manifest stability.

We believe that networks of string theories may provide a useful arena in which to explore new aspects of non-supersymmetric string dynamics and, potentially, novel mechanisms for the emergence and stabilization of string vacua. The present work is intended as a first step in that direction

The paper is organized as follows. In section \ref{sec:review} we review the worldsheet construction in \cite{Tachikawa:2026top} of the basic trivalent junction. In section \ref{sec:networks-gluing} we generalize the construction to effectively glue junctions and produce networks of 10d heterotic theories. Section \ref{sec:conjugate} introduces conjugate junctions,  which are glued in section \ref{sec:gluing} to build simple bubble networks, which we generalize in section \ref{sec:more-bubbles}. Section \ref{sec:transition} discusses transitions resolving such networks into disjoint of trivial configurations, and in section \ref{sec:stability} we consider possible avenues for the resolution of this instability. In section \ref{sec:compact} we describe compact networks. 

In section \ref{sec:graphs} we provide a graph theoretical description of the network topologies. Section \ref{sec:graph-rules} provides the graph rules, and section \ref{sec:bubbles-revisited} revisits the construction of bubble networks. In section \ref{sec:graph-classification} we build a systematic characterization of general graphs, and use it in section \ref{sec:graph-algebraic} to realize them in terms of $(0,1)$ heterotic worldsheets, illustrating it with several examples in section \ref{sec:graph-examples}. In section \ref{sec:sheets} we study a class of networks admitting a worldsheet description simpler than the one provided by our general algorithm. In section \ref{sec:x-sheets} we introduce the notion of sheets to encode a repeated pattern of edges, which we characterize in graph-theoretical terms in section \ref{sec:graph-multiple-sheet}, and continue exploring in section \ref{sec:xtilde-sheets}. In section \ref{sec:applications} we consider further applications of networks, including higher-dimensional networks (section \ref{sec:higher}), or compact networks as novel compactifications (section \ref{sec:compactifications}). We offer some final remarks in section \ref{sec:conclusions}.

\section{Trivalent junction of 10d non-tachyonic heterotic theories}
\label{sec:review}

In this section we describe the junction of the three 10d non-tachyonic heterotic string theories, proposed in \cite{Altavista:2026edv}, in particular via the construction of its $(0,1)$ heterotic worldsheet description \cite{Tachikawa:2026top}. 
 
The construction in \cite{Tachikawa:2026top} is based on the realization of the left-moving sector of the three heterotic $(0,1)$ worldsheet theories as a bosonic CFT $T$, admitting a non-anomalous $\IZ_2$ symmetry, the quotient bosonic CFT $T/\IZ_2$, and the quotient of a suitable fermionic CFT $(T\times q)/\IZ_2$, with $q$ is a certain $\IZ_2$-symmetric spin invertible theory. Although the construction in \cite{Tachikawa:2026top} holds for any $T$ and any symmetry $\IZ_2$, we focus on its application to the junction of the three heterotic theories\footnote{It should be clear that all results in this paper generalize straightforwardly to networks based on trivalent junctions of general theories $T$, $T/\IZ_2$ and $(T\times q)/\IZ_2$.\label{foot:general-junction}}. In this context, $T$ is the worldsheet CFT of the $E_8\times E_8$ theory, $T/\IZ_2$ is that of the $Spin(32)/\IZ_2$ theory, and $(T\times q)/\IZ_2$ describes the $SO(16)\times SO(16)$ heterotic theory. 

The junction construction is the following. We introduce the extra $(0,1)$ chiral multiplets $X$, ${\tilde X}$, and two Fermi multiplets $\Lambda,\,\tilde{\Lambda}$, with untilded or tilded fields being even or odd under the above mentioned $\IZ_2$, which is furthermore gauged. We introduce the $\IZ_2$-invariant $(0,1)$ superpotential interaction
\beqa
\int d\theta \Lambda({\tilde X}^2-X^2-Z)+\int d\theta {\tilde \Lambda} X{\tilde X}\, .
\label{tachikawa-supo}
\eeqa
where $Z$ is the 2d $(0,1)$ chiral multiplet describing one of the 10d coordinates. The corresponding scalar potential is
\beqa
V=({\tilde X}^2-X^2-Z)^2+(X{\tilde X})^2.
\eeqa
The minima are determined by
\beqa
X{\tilde X}=0\quad ,\quad {\tilde X}^2-X^2=Z.
\eeqa
The first condition means that at most only one out of $X,{\tilde X}$ can get a non-zero vev. Then, the solution of the second condition involves two possible regimes: 

$\bullet$ At $Z\gg 0$, we must have ${\tilde X}\neq 0$, hence $X=0$. There are two branches ${\tilde X}=\pm \sqrt{Z}$, but the $\IZ_2$ exchanges them ($\tilde{X}$ is $\IZ_2$-odd), so we keep one of them and need not impose any projection. The fields $X$ and the combination of ${\tilde X}$ and $Z$ orthogonal to the vacuum manifold are massive, and so are their fermions.  The massive fermions get the same mass (via fermion interaction $\lambda\psi_{\tilde{X}}\langle \tilde{X}\rangle+\tilde{\lambda}\psi_{X}\langle \tilde{X}\rangle$, with hopefully obvious notation). This branch hence supports a single copy of the theory $T$, namely a 10d $E_8\times E_8$ heterotic theory. 

$\bullet$ At $Z\ll 0$, we must have $X\neq 0$, hence ${\tilde X}=0$. There are two branches $X=\pm \sqrt{-Z}$, which are each mapped to itself by the $\IZ_2$ action ($X$ is $\IZ_2$-even). One may think that each of these branches supports a theory $T/\IZ_2$, but there is a subtlety. Upon integrating out the fermions, the mass of the Majorana fermion in ${\tilde \Lambda}$, ${\tilde X}$ has opposite sign  in the $X>0$ and $X<0$ branches (with fermion interaction $\lambda\psi_X\langle X\rangle+\tilde{\lambda}\psi_{\tilde{X}}\langle X\rangle$). This implies that there is an Arf-like term between the two branches that reproduces precisely the theory $q$ (we refer to \cite{Tachikawa:2026top} for details). So in the $X=+\sqrt{-Z}$ brane we have the theory $T/\IZ_2$ and in the $X=-\sqrt{-Z}$ brane we have the theory $(T\times q)/\IZ_2$. Namely, one $Spin(32)/\IZ_2$ theory and one $SO(16)\times SO(16)$ theory.

\begin{figure}[htb]
\begin{center}
\includegraphics[scale=.35]{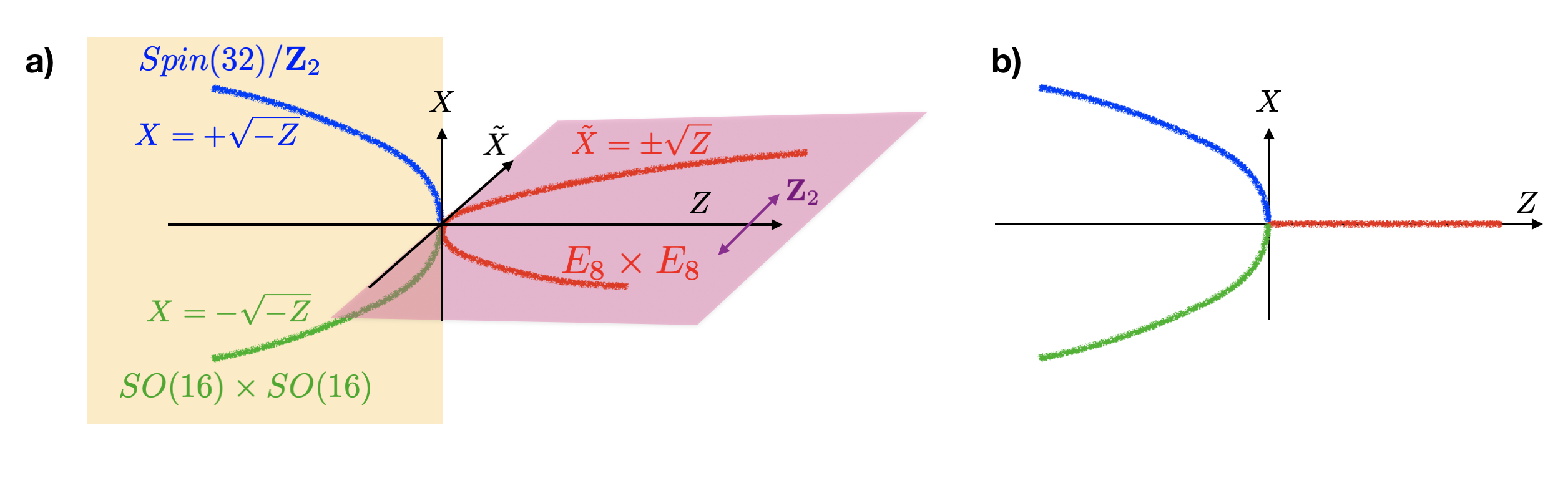}
\caption{\small a) Structure of the trivalent junction of non-tachyonic heterotic strings, in the description in \cite{Tachikawa:2026top}. b) A simplified description.}
\label{fig:heterotic-junction}
\end{center}
\end{figure}

The configuration is shown in Figure \ref{fig:heterotic-junction}a. For simplicity, in (b) we display only the $(X,Z)$ plane, and project the ${\tilde X}=\sqrt{Z}$ branch onto the $Z$ direction to simplify the depiction.

\section{Networks from gluing junctions}
\label{sec:networks-gluing}

In this section we show that a simple modification of the above setup already provides interesting examples of non-trivial networks, and consider some of their properties.

\subsection{Conjugate junctions}
\label{sec:conjugate}

\begin{figure}[htb]
\begin{center}
\includegraphics[scale=.45]{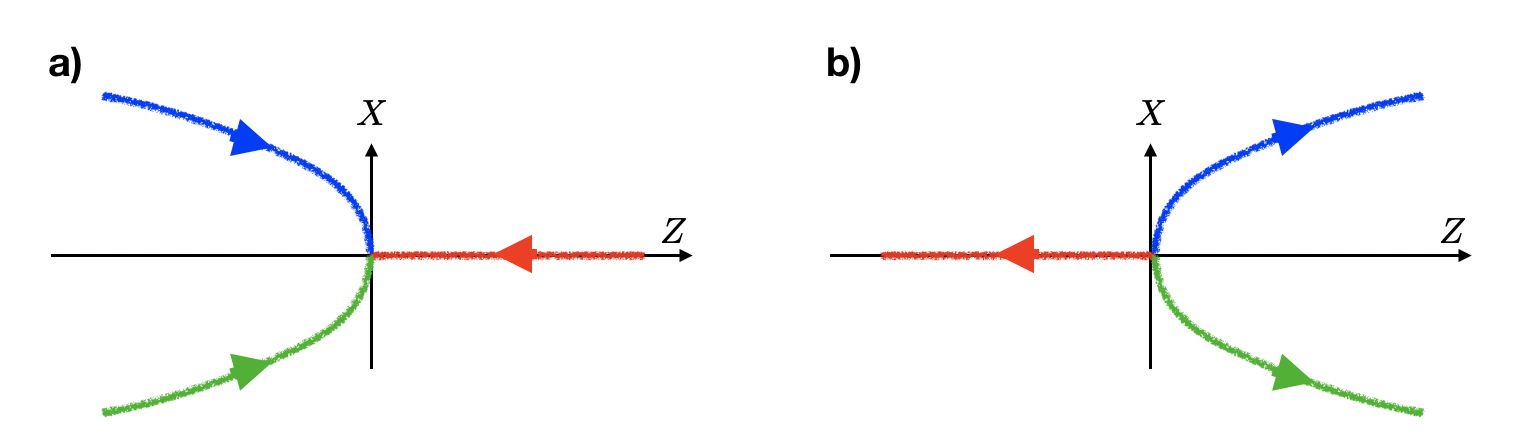}
\caption{\small a) The junction in Figure \ref{fig:heterotic-junction}b). b) The conjugate junction. The orientation of the branches is depicted to emphasize the key difference between both.}
\label{fig:two-junctions}
\end{center}
\end{figure}

Consider the same theory as above, but flip the sign of $Z$ in the superpotential, so that instead of (\ref{tachikawa-supo}), we have
\beqa
\int d\theta \Lambda({\tilde X}^2-X^2+Z)+\int d\theta {\tilde \Lambda} X{\tilde X}\, .
\label{tachikawa-supo-flipped}
\eeqa
The minima are determined by
\beqa
X{\tilde X}=0\quad ,\quad {\tilde X}^2-X^2=-Z\, .
\eeqa
When we have the same situation as above, but with a flip in the $Z$ coordinate, see Figure \ref{fig:two-junctions}, which gives the two variants of Figure \ref{fig:heterotic-junction}b. Although the two junctions could look identical, they differ in the orientation of the 10d heterotic string theories on their branches (which are chiral), so we refer to them as conjugate junctions. We depict these orientations in Figure \ref{fig:two-junctions} for clarity, although we will often leave it implicit.  

One important observation is that the relative location of the different branches with the positive or negative $Z$ direction is not significant. For example we can construct the same kind of junction of the three heterotic theories by taking
\beqa
\int d\theta \Lambda({\tilde X}^2\pm X+ Z)+\int d\theta {\tilde \Lambda} X{\tilde X}\, ,
\eeqa
The discussion for $X=0$ is the same as above, i.e. the two branches $\tilde{X}=\pm\sqrt{-Z}$ are exchanged with each other by the $\IZ_2$ action, resulting in one red branch supporting the theory $T$ located at $Z<0$. On the other hand, for ${\tilde{X}}=0$, we simply obtain the branch $X=\mp Z$, mapped to itself by the $\IZ_2$. The fermion mass ${\tilde \lambda}\psi_{\tilde X}$ now changes according to the sign of $Z$, from blue to green or viceversa, depending on the $\mp$ sign choice, so that for $Z>0$ we would have the theory $T/\IZ_2$, whereas for $Z<0$ we have $(T\times q)/\IZ_2$ (or viceversa). Hence, in this version of the junction the red branch at $Z<0$ is accompanied by a green or a blue branch. 

The punchline is that the topological class of the junction is only determined by the choice of overall orientation of the three participating theories, and is independent on the detailed directions in which the different theories stick out of it when embedded algebraically is a given set of coordinates.

\subsection{Gluing two junctions: Simple bubbles}
\label{sec:gluing}

\begin{figure}[htb]
\begin{center}
\includegraphics[scale=.38]{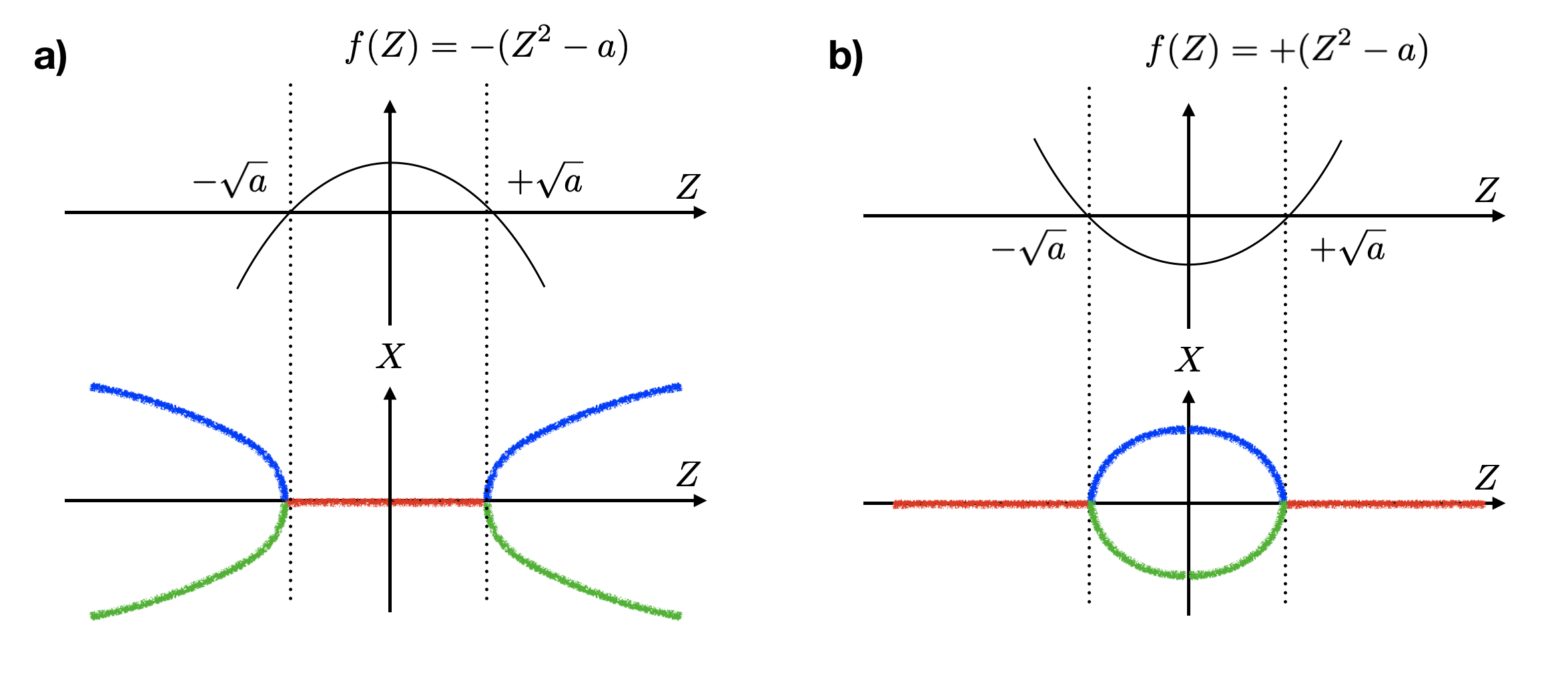}
\caption{\small a) The downward parabola and the resulting network, obtained by gluing a junction and its conjugate one. b) The upward parabola and the resulting network obtained by gluing the junction and the conjugate junction in a different way.}
\label{fig:network}
\end{center}
\end{figure}

It is straightfoward to glue together two conjugate junctions of the kind introduced above. Consider
\beqa
\int d\theta \Lambda({\tilde X}^2-X^2\mp (Z^2-a))+\int d\theta {\tilde \Lambda} X{\tilde X}\, ,
\label{general-superpotential-quadratic}
\eeqa
with $a$ some parameter $a>0$ (see later for other signs). Let us consider the two possible signs in turns:

$\bullet $ Consider first the case with the positive sign in (\ref{general-superpotential-quadratic}). As compared with section \ref{sec:review}, $Z$ is replaced by a function $f(Z)=- (Z^2-a)$. This is downwards concave parabola, with two zeros at $Z=\pm\sqrt{a}$, with positive slope at $Z=-\sqrt{a}$, and with negative slope at $Z=+\sqrt{a}$. This means that the theory describes a combination of the two junctions as shown in Figure \ref{fig:network}a. 

$\bullet $ Consider now the case with the negative sign in (\ref{general-superpotential-quadratic}). As compared with section \ref{sec:review}, $Z$ is replaced by a function $f(Z)=+ (Z^2-a)$, an upwards concave parabola, with a zero at $Z=-\sqrt{a}$ with negative slope and a zero at $Z=+\sqrt{a}$ with positive slope. This means that the theory describes a combination of the two junctions as shown in Figure \ref{fig:network}b. 

\begin{figure}[htb]
\begin{center}
\includegraphics[scale=.45]{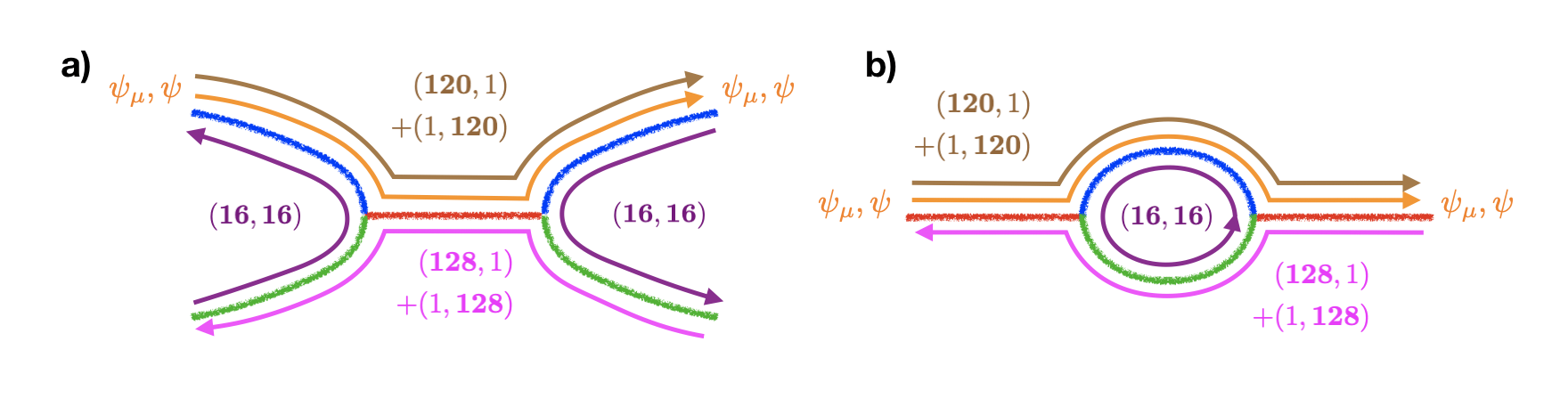}
\caption{\small Figures (a) and (b) give the chiral flow in the networks of Figure \ref{fig:network}a and b, respectively.}
\label{fig:chiral-flow}
\end{center}
\end{figure}

It is easy to see, by iterating the chiral flow in Figure \ref{fig:bouquet-chiral-flow} with opposite orientations in the conjugate junctions, that the chiral fields of the different branches propagate consistently across the network, see Figure \ref{fig:chiral-flow}. The same will hold for the more general networks discussed later, and it will receive a simple graph-theoretical description in section \ref{sec:graphs}.

One important observation is that the above configurations are not topologically protected against `decay'. Namely, even if you fix the asymptotics of the configuration, it is possible to recombine the zeroes of $f(Z)$ and have the two conjugate junctions annihilate each other, leaving a trivial configuration. In worldsheet language, there may be a non-trivial running of the parameter $a$ with the RG flow which leads to this kind of behaviour. We postpone its discussion until section \ref{sec:stability}. On the other hand, the effects of changing $a$, regarding this as an external parameter which can be tuned (rather than a quantity that runs dynamically with the RG flow), are discussed in section \ref{sec:transition}.

\subsection{More bubbles}
\label{sec:more-bubbles}

It is now straightforward to consider more general possibilities for $f(Z)$ and to derive the corresponding networks. In this simple setup, one just gets concatenations of the above ingredients in a linear sequence, according to the structure of zeroes of $f(Z)$ and the sign of its first derivative at those points.

\begin{figure}[htb]
\begin{center}
\includegraphics[scale=.45]{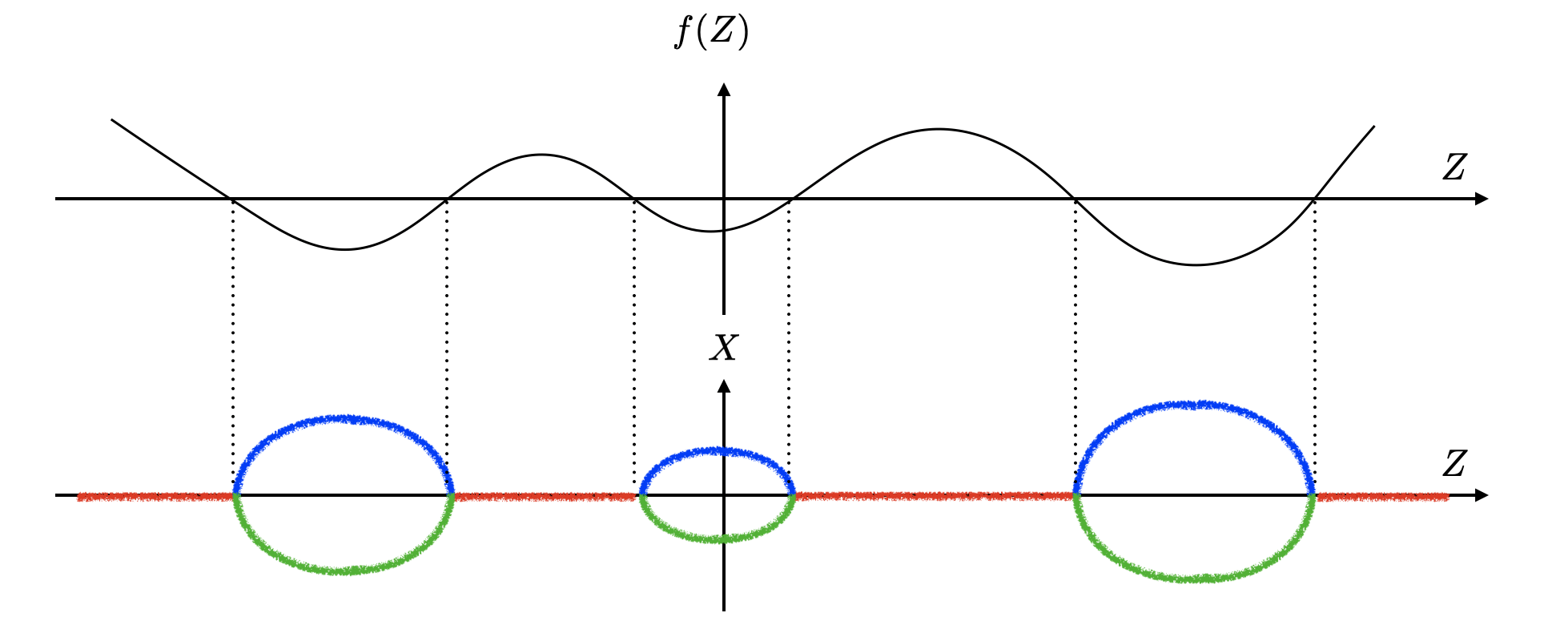}
\caption{\small An example of a network with a linear sequence of bubbles.}
\label{fig:more-bubbles}
\end{center}
\end{figure}

The simple junctions in sections \ref{sec:review} and \ref{sec:conjugate} arise from a superpotential in which the linear dependence in $Z$ can be considered just as the linear approximation in an expansion in $Z$ around one of the zeroes of the function. In fact, we already exploited this in section \ref{sec:gluing}, considering a quadratic function $f(Z)$ with two zeroes with opposite sign slopes. We can easily generalize to any general function
\beqa
\int d\theta \Lambda({\tilde X}^2-X^2-f(Z))+\int d\theta {\tilde \Lambda} X{\tilde X}\, .
\label{general-superpotential}
\eeqa
The vacuum equations are then
\beqa
X{\tilde X}=0\quad ,\quad {\tilde X}^2-X^2=f(Z)\, .
\label{general-vacuum}
\eeqa
Namely, near a zero at e.g. $Z=0$ we have to linear order $f(Z)\sim \pm Z$ and upon rescaling to reabsorb the coefficient we can reproduce the local behavior of basic junctions (with the sign of the slope determining which of two conjugate junctions we obtain). In general we can consider a general function $f(Z)$ and, near each zero $f(z_0)=0$, we will have a junction (if the first derivative $f'(z_0)>0$) or the conjugate junction (if the first derivative $f'(z_0)<0$). This allows to glue junctions to form more general networks, as illustrated in Figure \ref{fig:more-bubbles}.

\subsection{The transition}
\label{sec:transition}

\begin{figure}[htb]
\begin{center}
\includegraphics[scale=.35]{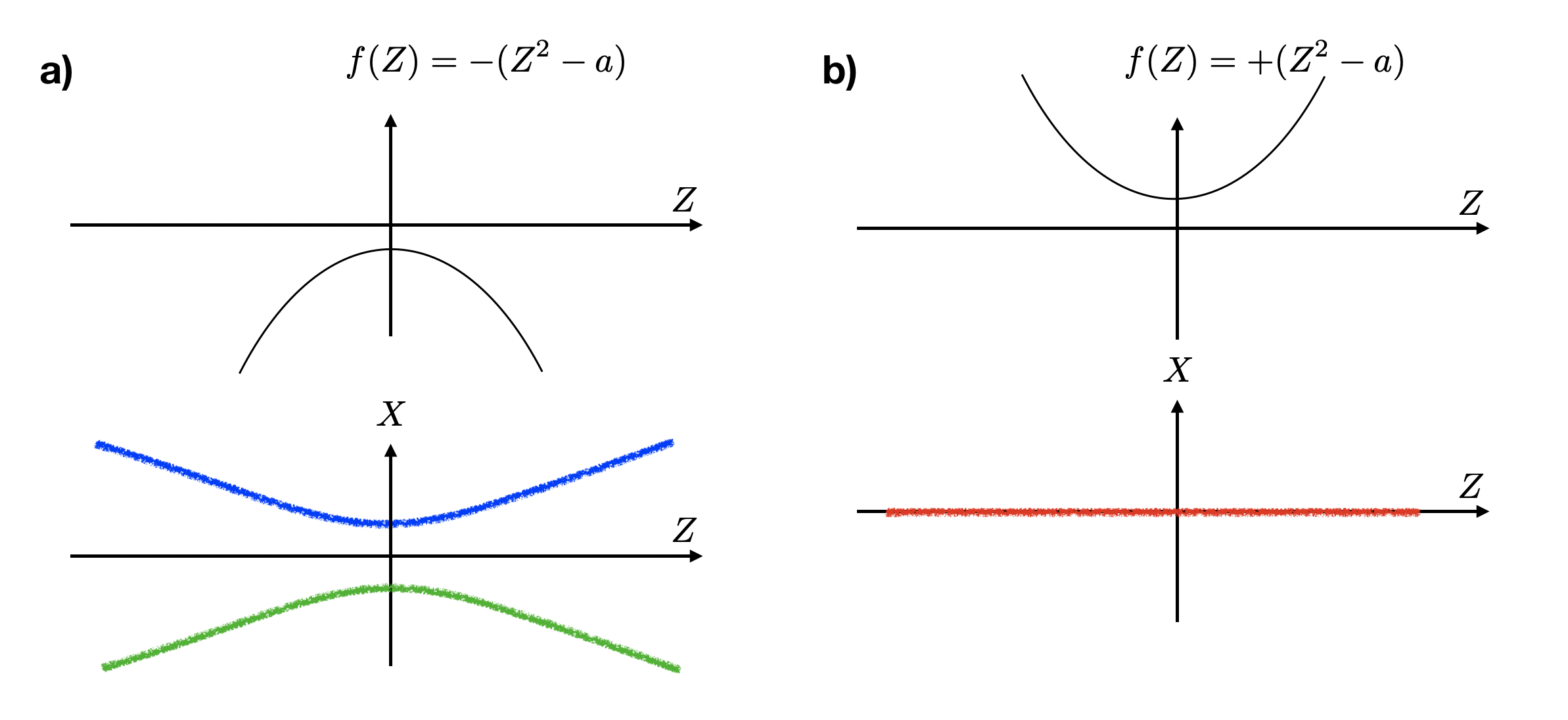}
\caption{\small a) The trivial version of the network in Figure \ref{fig:network}a. b) The trivial version of the network in Figure \ref{fig:network}b.}
\label{fig:trivial-network}
\end{center}
\end{figure}

Let us now go back to the original construction in section \ref{sec:gluing}, and see the effect of changing the sign of $a$. In particular let us consider
\beqa
f(Z)=\pm (Z^2-a)\; , \quad {\rm for}\; a<0\, .
\label{parabola2}
\eeqa
Now the function $f(Z)$ does not have any zero, its value has a fixed sign. For the negative sign in (\ref{parabola2}), $f(Z)<0$ for any $Z$, so the vacuum equations (\ref{general-vacuum}) require ${\tilde X}=0$, $X=\pm\sqrt{-f(Z)}$. We get two disconnected blue and green lines, namely the 10d $Spin(32)/\IZ_2$ and $SO(16)\times SO(16)$ theories, for any $Z$. The former non-trivial network in Figure \ref{fig:network}a has trivialized into the situation shown in Figure \ref{fig:trivial-network}a. For the positive sign case in (\ref{parabola2}), $f(Z)>0$ for any $Z$, and the vacuum equations require $X=0$ and $\tilde{X}=\pm \sqrt{f(Z)}$. Both branches are identified by the $\IZ_2$, so we obtain a single one supporting the 10d $E_8 \times E_8$ theory. This is depicted as a single red line in Figure \ref{fig:trivial-network}b, and represents the trivialized version of the network in Figure \ref{fig:network}b. 

\begin{figure}[htb]
\begin{center}
\includegraphics[scale=.35]{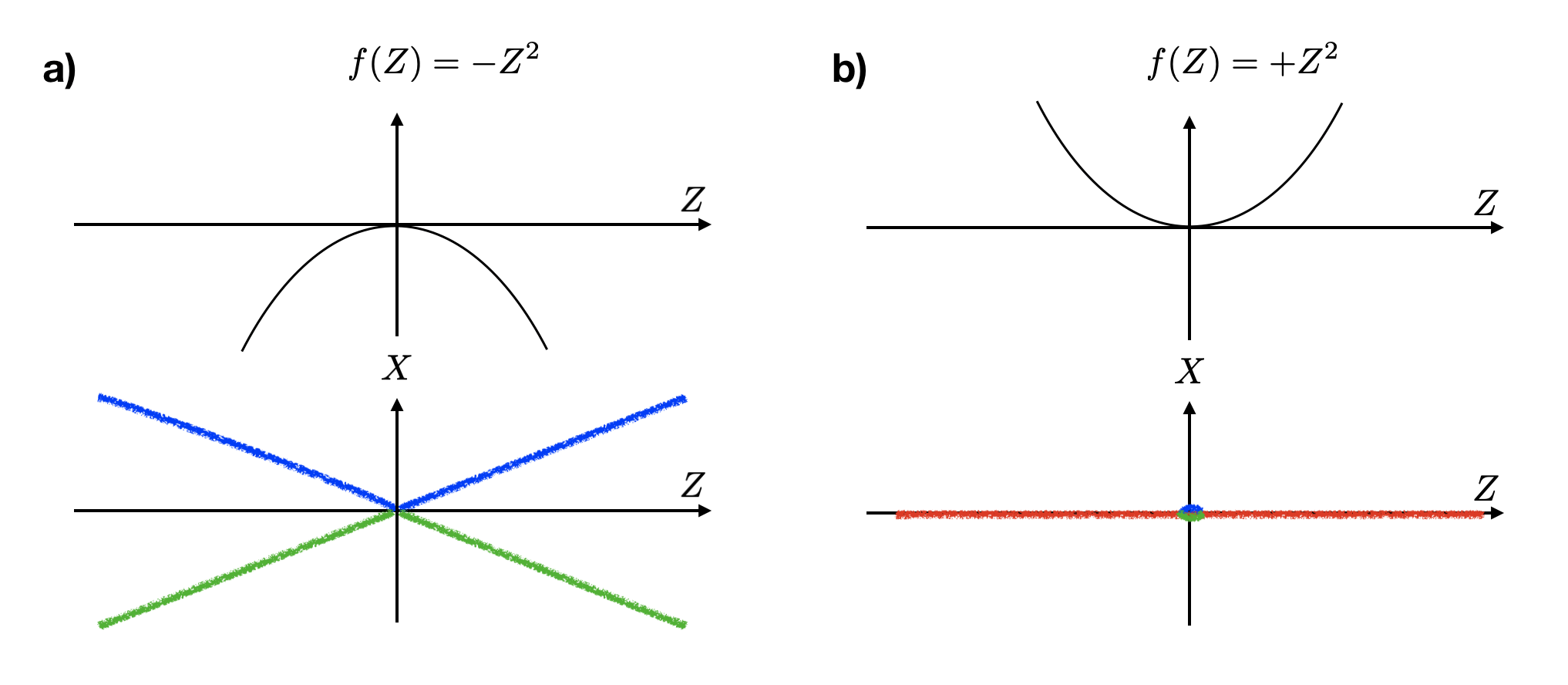}
\caption{\small a) The critical version of the network in Figures \ref{fig:network}a, \ref{fig:trivial-network}a. b) The critical version of the network in Figures \ref{fig:network}b, \ref{fig:trivial-network}b. The figure is actually an $X$-shaped image, similar to figure (a), but in the $({\tilde X},Z)$-plane, recall Figure \ref{fig:heterotic-junction}a.}
\label{fig:critical-network}
\end{center}
\end{figure}

Our meaning of a network becoming trivial is made more clearly by considering the critical situation between the two cases $a>0$ and $a<0$, namely $a=0$, which are depicted in Figures \ref{fig:critical-network}. Note that, for the case $f(Z)<0$, this can be regarded as a 4-valent junction, of two 10d $Spin(32)/ \mathbf{Z}_2$ and two $SO(16)\times SO(16)$ heterotic theories. It describes two theories defined on two 10d flat spacetimes, which intersect over a 9d subspace, without exchange of any chiral matter.  We will call such junctions non-chiral, because their chiral flow is trivial. Hence, we obtain that a non-chiral junction of four lines with pairwise related colors can be decomposed into two trivalent junctions. It is similarly possible to generalize and construct non-chiral intersections of more 10d heterotic theories, of different kinds, but they can always be regarded as limits of networks built using the basic trivalent junction (and its conjugate). We will discuss this more in section \ref{sec:graphs}.

\begin{figure}[htb]
\begin{center}
\includegraphics[scale=.35]{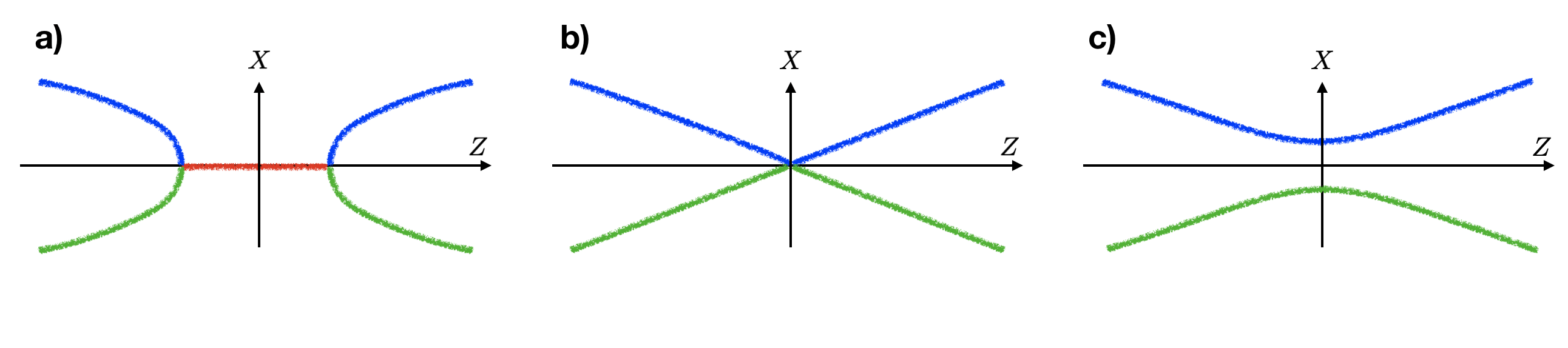}
\caption{\small The network for $f(Z)=-(Z^2-a)$ in the transition $a>0$ (figure (a)) to $a=0$ (figure (b)) to $a<0$ (figure (c)).}
\label{fig:transition}
\end{center}
\end{figure}

\begin{figure}[htb]
\begin{center}
\includegraphics[scale=.35]{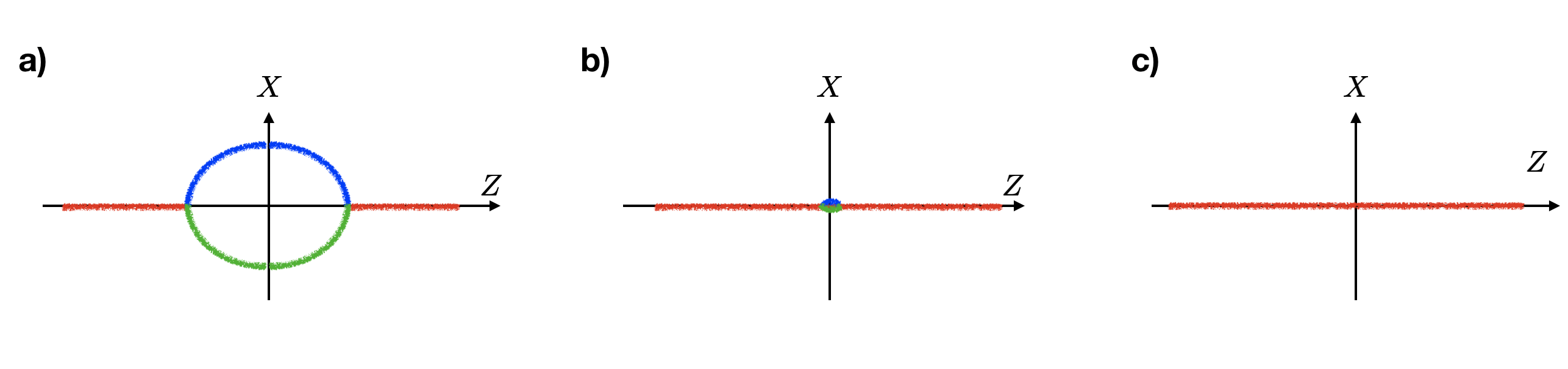}
\caption{\small The network for $f(Z)=+(Z^2-a)$ in the transition $a>0$ (figure (a)) to $a=0$ (figure (b)) to $a<0$ (figure (c)). Actually this Figure and Figure \ref{fig:transition} would really look very similar, if we had depicted the direction ${\tilde X}$.}
\label{fig:transition2}
\end{center}
\end{figure}

We can in fact study the transition from $a>0$ to $a<0$ across $a=0$, as shown in Figure \ref{fig:transition} for the case of $f(Z)=-(Z^2-a)$, and Figure \ref{fig:transition2} for $f(Z)=+(Z^2-a)$. We would like to note that  these two Figures would actually be very similar if we had depicted the direction ${\tilde X}$. For instance, the red line in Figure \ref{fig:transition}a is really an ellipse in the $({\tilde X},Z)$-plane, just like the green/blue ellipse in the $(X,Z)$ plane in Figure \ref{fig:transition2}a. Similarly, the red lines in \ref{fig:transition2}b form an $X$-shaped structure in the $({\tilde X},Z)$-plane, just like the green/blue one in Figure \ref{fig:transition}b. Finally,  the red line in \ref{fig:transition2}c is given by two disjoint lines in the $({\tilde X},Z)$-plane, just like the green/blue ones in Figure \ref{fig:transition}c. The $\IZ_2$ quotient projects down these shapes, so that the red ellipse becomes the red interval in Figure \ref{fig:transition}a, and the $X$-shaped or disjoint red lines become the single red line in Figures \ref{fig:transition2}b, c.

The above diagrams encode a fascinating phenomenon: as the parameter $a$ is changed, the two disjoint theories in Figure \ref{fig:transition}c can cross as in Figure \ref{fig:transition}b and create a new stretched branch of a 10d string theory, Figure \ref{fig:transition}c. This is a remarkable phenomenon, to our knowledge not considered in the literature. The phenomenon can be regarded as reminiscent of Hanany-Witten brane creation effects \cite{Hanany:1996ie} (but now for whole 10d theories) or to flop transitions in CY geometries (but now for whole 10d geometries). Also, regarding Figure \ref{fig:transition2} the analogous process of moving from Figure \ref{fig:transition2}c, to b and to a, corresponds to starting from a single 10d theory and nucleating a bubble of two simultaneous theories which are disconnected except at the boundary of the bubble. This is again a novel kind of phenomenon to the best of our knowledge not considered in the literature.

\subsection{Stability}
\label{sec:stability}

The trivalent junction is topologically protected by the chiral flow it supports. On the other hand, more complicated networks, starting from the bubble diagrams of section \ref{sec:gluing}, do not enjoy a similar protection. Namely, even if two networks are topologically different as graphs, they may not be protected against physical processes which e.g. remove non-trivial loops of the network, but leave the chiral flows outside that local region unchanged. For the bubble diagrams, this is manifest for instance from Figure \ref{fig:chiral-flow}, where the chiral flows in the presence of bubbles allow for contracting the bubble to a point and removing it, leaving either one or two disconnected 10d heterotic string theories. This decay process is precisely that mentioned in the previous section, c.f. Figures \ref{fig:transition} and \ref{fig:transition2}.  

This implies that, even if one formulates a worldsheet theory describing the bubble configuration, the RG flow to the infrared CFT may erase the presence of the bubble completely. On general grounds, we may view the process as the annihilation of two conjugate junctions, and therefore we expect it to be occur generically, although we do not have a quantitative computation of its precise dynamics. We may also expect a similar behavior in more involved networks, which may in general have several topologically allowed decay channels.
One may then question the possible usefulness of discussing such network configurations, even more so in the more involved examples in later sections. In this section we would like to propose several reasons to support the interest in this analysis.

The first is that it may be possible to consider spacetime-dependent configurations in which the decay of a network into a simpler one is described as an on-shell process, such as a time-dependent description of the decay, or a space-dependent configuration in which such process occurs as we move in some spatial direction. This kind of configuration is closely related to the higher-dimensional networks considered in section \ref{sec:higher}, which builds on the understanding of off-shell configurations as a first step.

A second motivation is that it is often interesting to study dynamically unstable configurations as a preliminary step to learn about their possible stabilization mechanisms. An analogy is the study of compactifications on spheres (which lead to unstable configurations due to the radion potential triggering the collapse of the sphere) and their subsequent stabilization via introduction of fluxes.

Furthermore, it is oftentimes useful to consider unstable configurations for their own sake, even in the absence of stabilization mechanisms, as a tool to enlarge the configuration space of a theory. An analogy is the use of  brane-antibrane pairs, which despite their tachyonic instability allow for a much deeper understanding of RR charges and fields in terms of K-theory, and of lower-dimensional branes as tachyon solitons \cite{Witten:1998cd,Sen:1999mg}.

Finally, there may be even more fundamental reasons supporting the physical interest of our network configurations. There are recent proposals of compactifications of M-theory and string theories on quantum singular spaces such as $\IS^1\vee \IS^1$ \cite{Baykara:2026gem,Altavista:2026evd,Baykara:2026vdc,Altavista:2026brr}, which may admit a definition in terms of (a suitable quantum version of) junction of 10d string theories, as recently explored in \cite{Basile:2026trt}. As we will further discuss in section \ref{sec:compactifications}, our networks share some intriguing features with such configurations. For instance, the fact that different fields may obey different boundary conditions on networks (including that they propagate only on different subsets of the network), as a consequence of the different junctions conditions, e.g. those dictated by the chiral flows. Hence, one may view the exploration of networks of junctions 10d theories in the context of the incipient exploration of quantum singular spaces, and their role in string duality. 

Hoping that these motivations suffice to convince the reader of the interest of these configurations, we continue considering further properties of networks, and their more systematic characterization and construction in terms of (admittedly off-shell) heterotic worldsheet theories.

\subsection{Compact networks}
\label{sec:compact}

\begin{figure}[htb]
\begin{center}
\includegraphics[scale=.5]{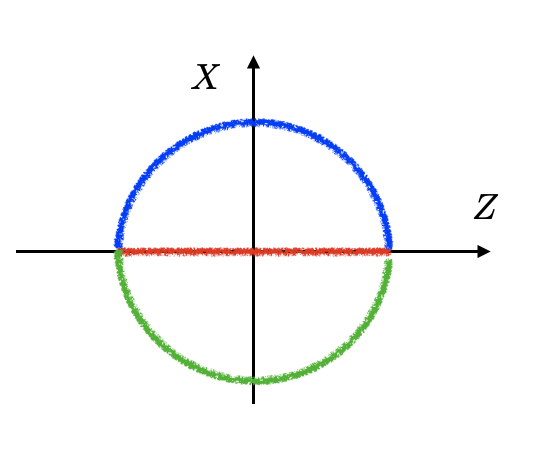}
\caption{\small The simplest compact network, with two junctions and one copy of each of the three theories. The picture in the $\IZ_2$ covering space is obtained by sticking out an additional direction ${\tilde X}$ transverse to the figure, and promoting the red segment to a full circle in the $(Z,{\tilde X})$-plane.}
\label{fig:compact-network}
\end{center}
\end{figure}

It is easy to use the above ideas to build compact networks, namely those in which all lines are finite. The simplest such network contains two junctions and three lines, one of each color, as shown in Figure \ref{fig:compact-network}. The superpotential is similar to those above, but flipping the sign of $X^2$ to remove the non-compact directions of the previous cases, namely
\beqa
W=\Lambda({\tilde X}^2+X^2+Z^2-a)+{\tilde \Lambda}X{\tilde X}\;, {\rm for}\; a>0\, .
\eeqa

This configuration provides a good illustration of the statement in section \ref{sec:conjugate} that the junction specifies a topological gluing of the three theories, which can receive different interpretations according to how the junction is read. The junctions in the compact configuration are identical to those in the previous section, with a sign flip of the $X^2$ term which makes all theories arise `on the same side' of the junction in the $Z$ coordinate.

Finally, this configuration can be regarded as a compactification to 9d of a combination of the three non-tachyonic heterotic strings. This viewpoint will be further explored in section \ref{sec:compactifications}.

The generalization of the above model to more general functional dependences on $Z$, namely
\beqa
W=\Lambda({\tilde X}^2+X^2+f(Z))+{\tilde \Lambda}X{\tilde X}\;, {\rm for}\; a>0\, ,
\eeqa
lead to disconnected copies of the basic compact diagram, stretching between consecutive pairs of zeros of the function $f(Z)$.

\medskip

It is straightforward to construct further examples of networks, which we leave as an exercise for the interested reader. Rather we move to the formalization of the above ingredients, which will provide a systematic classification of networks, and their explicit construction in terms of $(0,1)$ heterotic worldsheet theories.

\section{Networks as graphs}
\label{sec:graphs}

We have shown that there exist networks of 10d string theories with branches supporting the non-tachyonic heterotic strings, and with trivalent junctions involving each of these theories, with specific orientations so as to achieve a consistent chiral flow. We now develop a graph theoretical formulation of the topology\footnote{We clarify that the meaning of topological equivalence in this section is without allowing for annihilation of conjugate junctions.} of the resulting general networks. This will also provide the tools to realize the networks in terms of explicit $(0,1)$ worldsheet theories, to recover the previous examples, and to vastly generalize them.

\subsection{The graph rules}
\label{sec:graph-rules}

We now describe the topology of the networks in terms of a mathematical graph, namely a set of edges joining at a set of vertices. For our graphs there is no notion of faces, and in general we will not require that the graph is embedded in any ambient space. On the other hand, we must allow for three different kinds of edges (we add a coloring), endow them with an orientation, and allow them to join at only two kinds of (conjugate) trivalent vertices. This will result in a specific class of graphs, whose key features we now explain.

We will focus on graphs describing networks of the three 10d non-tachyonic heterotic strings, joining at trivalent junctions involving each of these theories with a consistent chiral flow. Consequently, each edge will correspond to a 10d heterotic theory, to which we assigne one out of three colors, say red, green and blue, as in the previous section. Edges are oriented, reflecting the fact that the 10d heterotics are chiral, and we denote the orientation with an arrow. Edges can be finite (they join two vertices) or semi-infinite (they are attached to one vertex and stick out to infinity at their other end).

\begin{figure}[htb]
\begin{center}
\includegraphics[scale=.4]{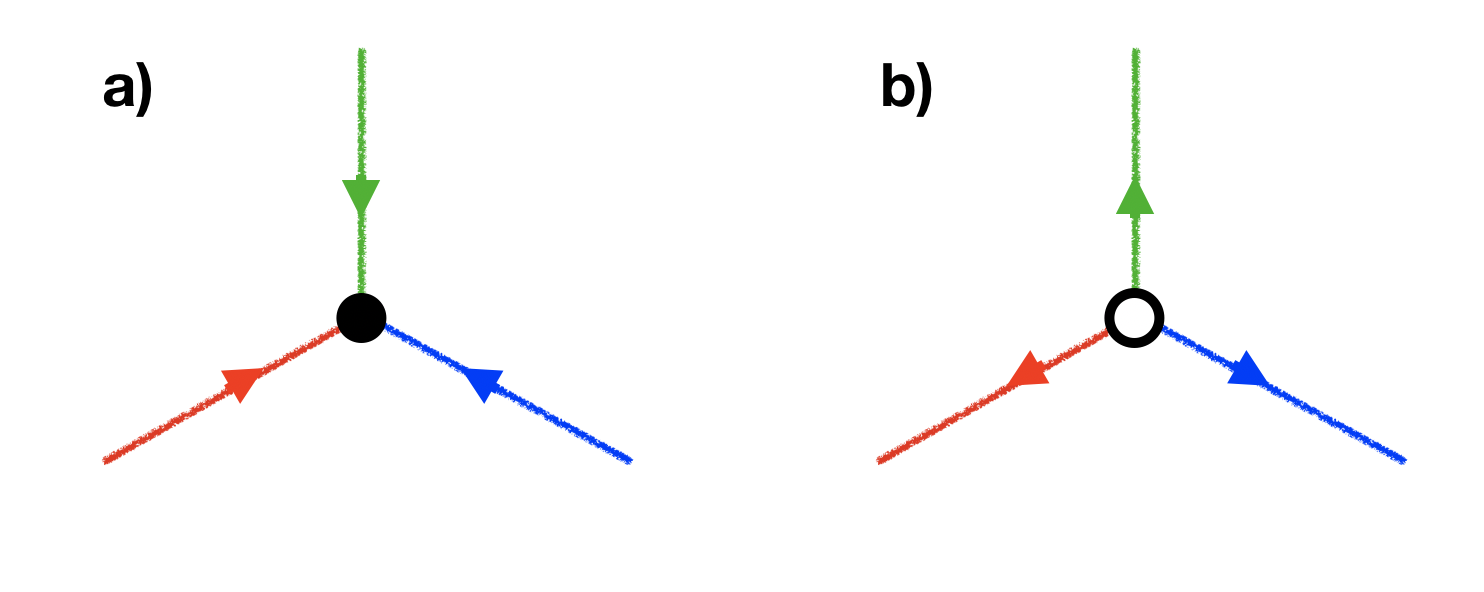}
\caption{\small The two vertices describing the junction of the three 10d non-tachyonic heterotic strings. The cyclic ordering of the edges is irrelevant.}
\label{fig:heterotic-vertices}
\end{center}
\end{figure}

Each vertex (or node) represents a trivalent junction of the three heterotic theories. There are only two kinds of vertices, one with all edges with incoming orientation (which we denote with as a black vertex, or a vertex of $+$ kind) and one with all edges with outgoing orientation (white vertex, or of $-$ kind).  They are related to flipping the chiralities of all fields in all the heterotic theories, an operation compatible with the chiral flow. The vertices and edges are shown in Figure \ref{fig:heterotic-vertices}. Note that the cyclic ordering of the edges is irrelevant.

\begin{figure}[htb]
\begin{center}
\includegraphics[scale=.4]{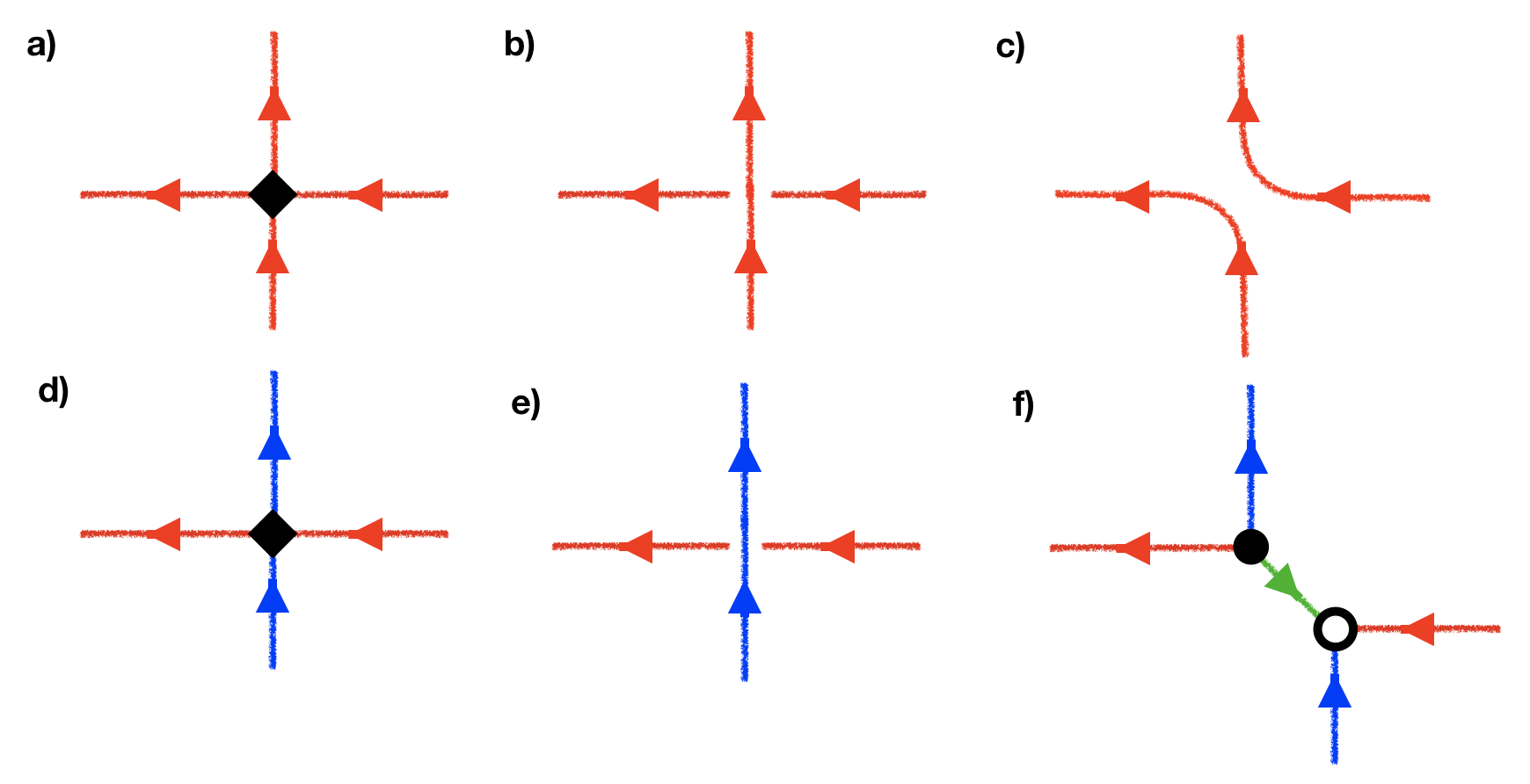}
\caption{\small Some examples of resolutions of non-chiral junctions. The 4-valent junction of same color edges (a) can be resolved into disjoint graphs, in two different ways, (b) and (c). The 4-valent junction of edges of two different colors (d) can be resolved into a disjoint graph (e) or two conjugate trivalent junctions (f).}
\label{fig:junctions-resolutions}
\end{center}
\end{figure}

We emphasize that we do not consider graphs with non-chiral junctions, e.g. 4-valent nodes with pair-wise related theories. As described in section \ref{sec:transition}, one can consider graphs in which for each incoming edge of one color there is one outgoing one of the same color. Such graphs have a trivial chiral flow (we call them non-chiral), so they are not protected against ``resolution''. If the incoming edges are of the same color, the junction can be smoothed out by recombining the edges of the same color and separating them so that they become disjoint, in either of two inequivalent possible ways. If the incoming edges are of two different colors, we can still do the recombination into non-intersecting edges as above, or we can resolve the junction by splitting it into two new conjugate trivalent vertices, joined by and edge of the third color. This is shown in Figure \ref{fig:junctions-resolutions}.  A similar comment applies to non-chiral junctions with higher valence.

It is easy to convince oneself that any consistent junction can be simplified to just a set of basic bouquets, possibly with some non-chiral junctions stacked on top. And that it is always possible to resolve the non-chiral junctions into either disjoint ones, or pairs of conjugate vertices. This means that any network involving these higher-valence junctions can be regarded as a limit of a network with only trivalent junctions. Hence, we will focus on constructing networks with only black and white trivalent nodes.

\medskip

Let us now draw some simple implications from the above ingredients. First note that because of the orientation of the edges, there are no edges joining vertices of the same color, implying that the graph is bipartite\footnote{Bipartite graphs (but with no coloring of edges, and with an embedding as tilings of $\IT^2$, or in general of Riemann surfaces) play an important role in the description of D3-branes at toric CY singularities (for reviews, see e.g. \cite{Kennaway:2007tq, Garcia-Etxebarria:2006ngz}). We note that however there is no obvious link between these two applications of graphs.}.

The set of edges of a given color, e.g. red, define a perfect matching of the bipartite graph. Namely, the complete set of vertices is separated in pairs, with the two vertices in each pair joined up by a single edge in the perfect matching. Hence, each of our graphs is a superposition of three perfect matchings (one per color). Interestingly, we note that a perfect matching may be regarded as a linear map between black and white vertices. In our case, it corresponds to the adjacency matrix (i.e. the $ij$ entry is $1$ if there is an edge between the vertices $i$ and $j$, and $0$ otherwise) associated to the edges of the corresponding color.

\begin{figure}[htb]
\begin{center}
\includegraphics[scale=.4]{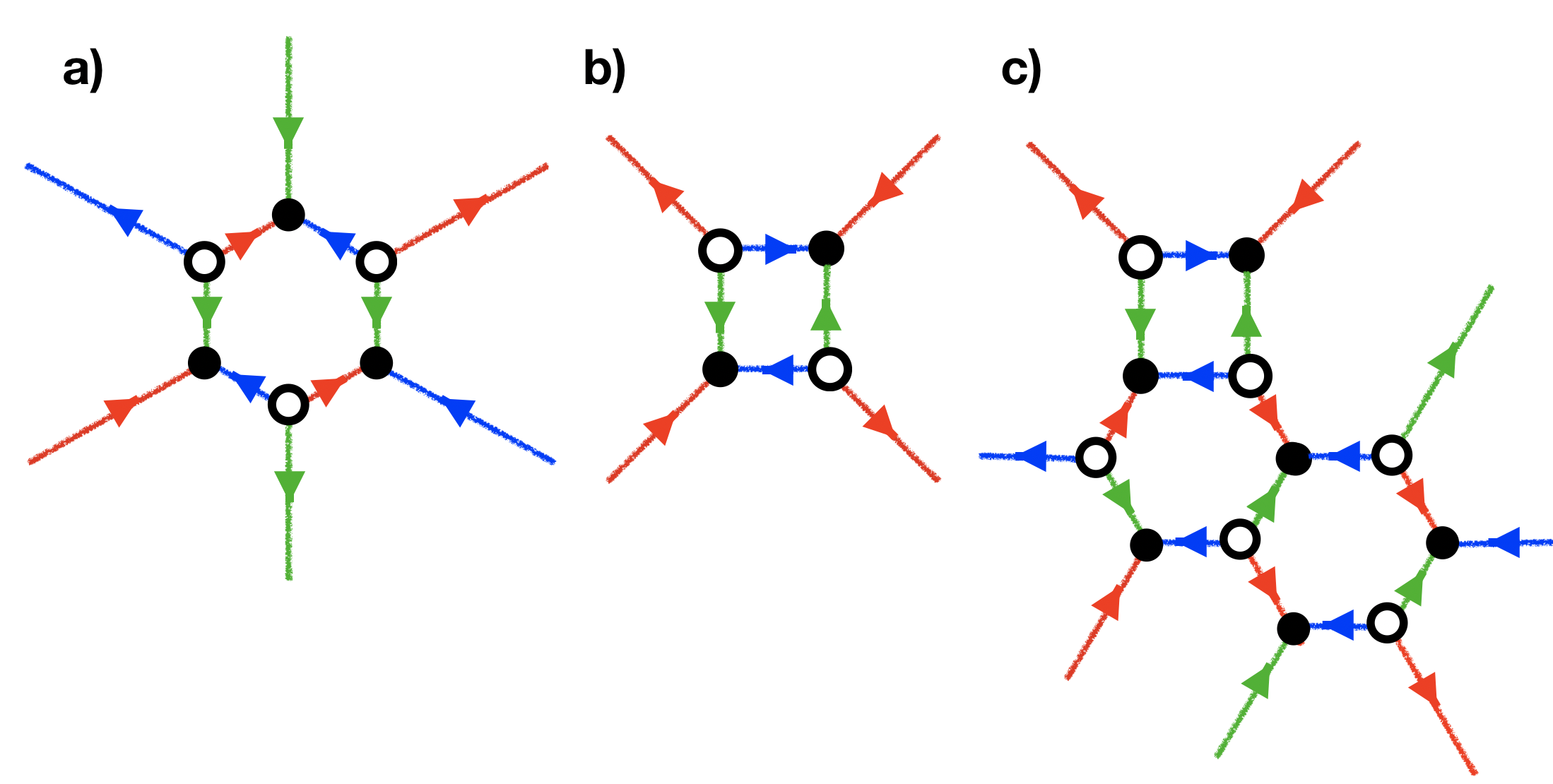}
\caption{\small Some examples of consistent graphs.}
\label{fig:new-dream-graphs}
\end{center}
\end{figure}

Finally, note that, since each edge (of any color) joins a white and a black node, it follows that any closed loop in the graph must have an even number of edges. This is already an interesting constraint that rules out many naive graphs, but it is easy to build consistent many graphs,  see e.g. Figure \ref{fig:new-dream-graphs}.

\begin{figure}[htb]
\begin{center}
\includegraphics[scale=.4]{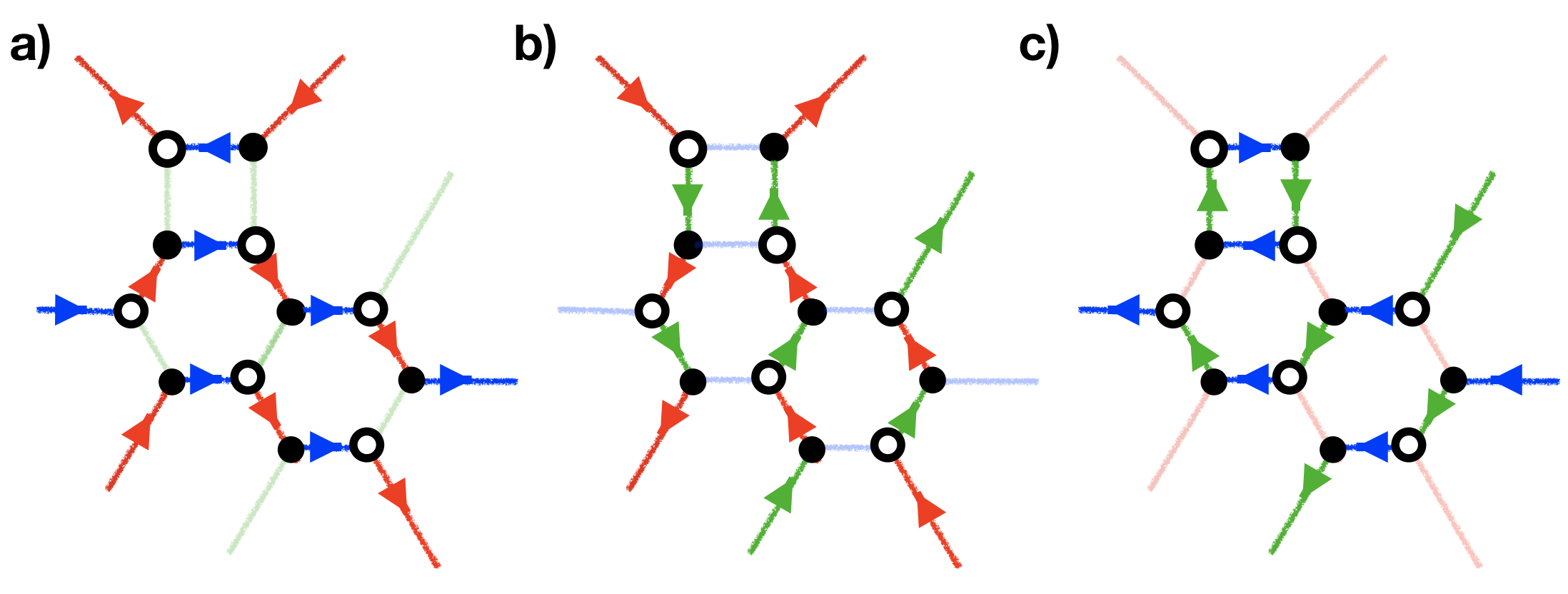}
\caption{\small The two-colored paths of the graph in Figure \ref{fig:new-dream-graphs}c. We show in semi-transparent lines the edges not participating in the path. (a) The paths $R{\ov B}$; (b) the paths $G{\ov R}$; (c) the paths $B{\ov G}$}
\label{fig:paths}
\end{center}
\end{figure}

In bipartite graphs one can form some special paths, as follows. Choose two colors, say red and blue, and superimpose the set of red edges and of blue edges, with orientation flipped for the latter, we denote them by $R{\ov B}$ paths. The corresponding combination of edges forms a set of oriented curves, which are either closed loops (if all edges in the curve are finite) or infinite curves (if there are semi-infinite incoming and outgoing edges). We will refer to them as two-colored paths\footnote{They have some analogy to the zig-zag paths of brane tilings, but in that case the notion of face (or cyclic ordering of edges in the vertex) is relevant, and in that case we do not have colored edges. Hence, probably there is not clear relation between both constructions of paths.}. Clearly one can form similar paths for red and green ($R{\ov G}$), or blue and green ($B{\ov G}$). Taking the pair in the reversed order simply flips the orientation of the path. So overall we get three sets of two-colored paths. In the heterotic context, it is easy to see that they correspond to the chiral flow of fields, c.f. Figure \ref{fig:bouquet-chiral-flow}. In Figure \ref{fig:paths} we give the two-colored paths for the graph in Figure \ref{fig:new-dream-graphs}c.

\begin{figure}[htb]
\begin{center}
\includegraphics[scale=.35]{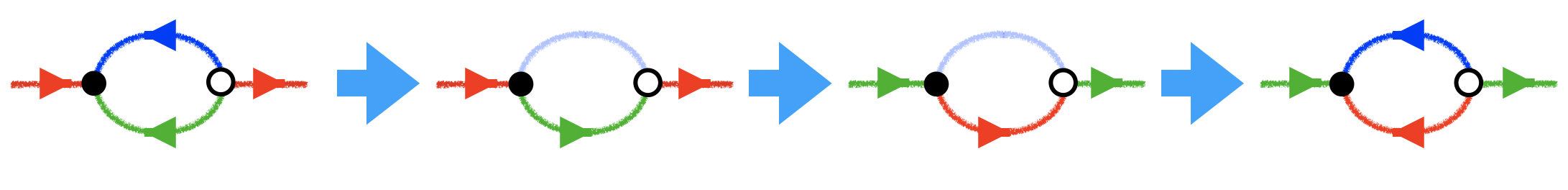}
\caption{\small Color flip of an $R{\ov G}$ path in a simple bubbling diagram. We start with the diagram, and isolate the $R{\ov G}$ path, which in this case has only one connected component. We then flip red and green, and then reconstruct the new diagram.}
\label{fig:path-color-flip1}
\end{center}
\end{figure}

\begin{figure}[htb]
\begin{center}
\includegraphics[scale=.4]{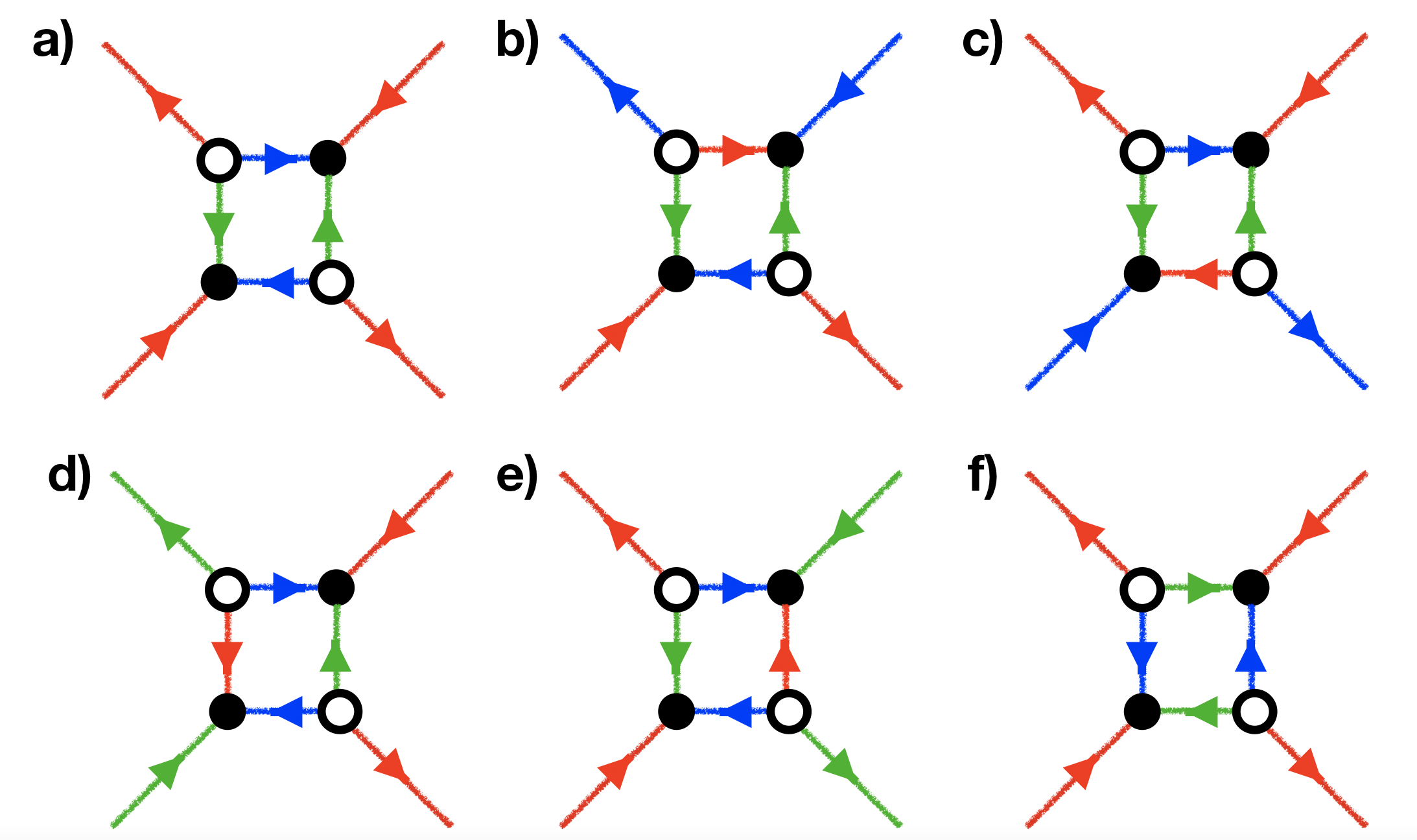}
\caption{\small Color flips of paths in the diagram in Figure \ref{fig:new-dream-graphs}b. (a) the original diagram; b) The diagram after color flip of the upper component of the $R{\ov B}$ path; c) color flip of the lower component of  the $R{\ov B}$ path; d) color flip of the left component of the $R{\ov G}$ path; e) color flip of the right component of the $R{\ov G}$ path; f) color flip of the only component of the $B{\ov G}$ path.}
\label{fig:path-color-flip2}
\end{center}
\end{figure}

Incidentally, we note that, starting with a graph, we can generate another (generically different) one as follows. Consider two colors, say red and green, and build the corresponding  $R{\ov G}$ path, which in general consists of several different connected components. Pick a particular connected component, and flip the red edges to green and viceversa (without touching any path in any other connected component, or any other color). This is shown in an example in Figure \ref{fig:path-color-flip1}. Clearly this works for any other choices of colors, etc. We will refer to this as the color flip of an $R{\ov G}$ path (component), and similarly for other color choices. For instance, in Figure \ref{fig:path-color-flip2} we show different graphs obtained from that in Figure \ref{fig:new-dream-graphs}b upon  color flips of different paths.
Clearly, the only change in the diagram is a different coloring of the edges.

Finally, it will be useful to introduce the notion of compact graph, as one with no semi-infinite edges. In compact graphs, there are some rules constraining the total numbers $V_\pm$ of vertices of each kind, and the number  $E_i$ of edges of each color, with $i=1,2,3$ corresponding to e.g. red, green, blue. For instance, there is exactly one edge of each color per white vertex, and similarly for black nodes, so we have $E_i=V_+=V_-$ for each $i=1,2,3$. Also, adding up the total numbers of edges $E=E_1+E_2+E_3$ and of vertices $V=V_++V_-$, they satisfy $2E=3V$. 

\begin{figure}[htb]
\begin{center}
\includegraphics[scale=.4]{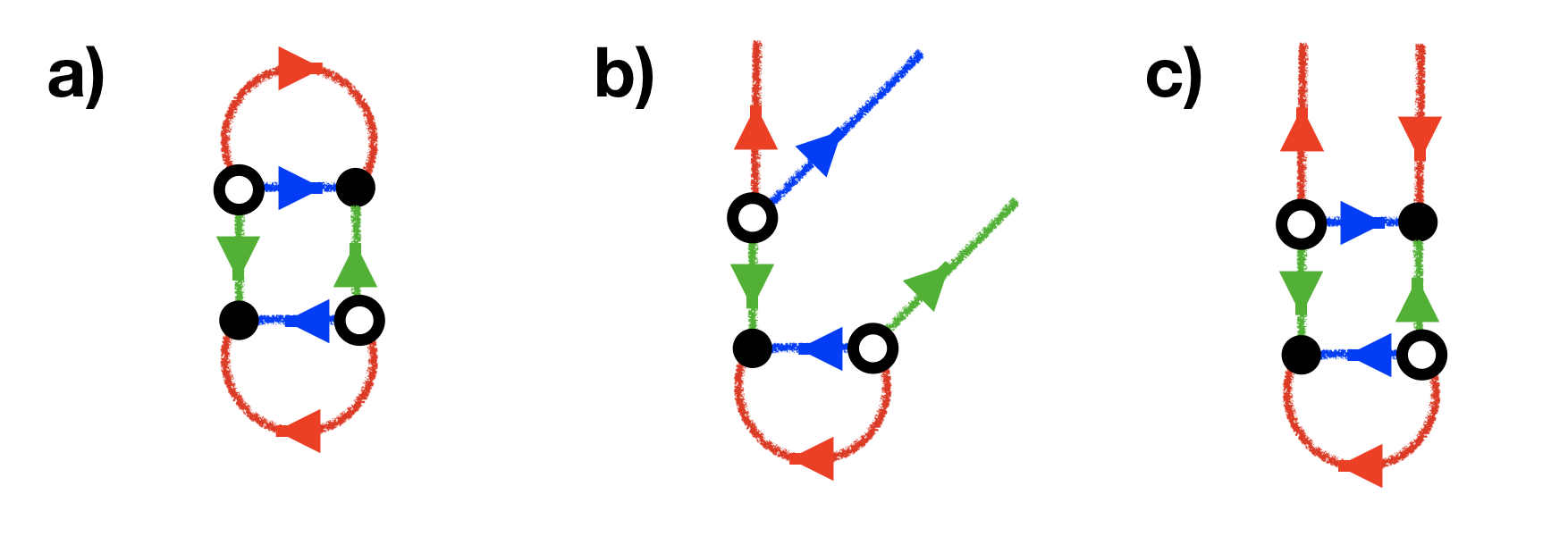}
\caption{\small An example of two ways to generate non-compact graphs from a compact one. a) The compact graph. b) Non-compact graph obtained by sending to infinity the upper-right black node. c)  Non-compact graph obtained by sending to infinity the middle region of the uppermost red edge. }
\label{fig:compact-noncompact}
\end{center}
\end{figure}

In general, we can also discuss non-compact graphs by regarding them as limits of compact ones, in which either some vertices are taken to infinity, or some finite edge is elongated exaggeratedly until its middle part is taken to infinity, see Figure \ref{fig:compact-noncompact}. It is easy to derive rules for the numbers of vertices and of edges of each color for general non-compact graphs, for instance $2E_i+L_i=V_++V_-$ (where $L_i$ is the number of external legs of the given color), but we will not make explicit use of them.  

\subsection{Bubbles revisited}
\label{sec:bubbles-revisited}

\begin{figure}[htb]
\begin{center}
\includegraphics[scale=.4]{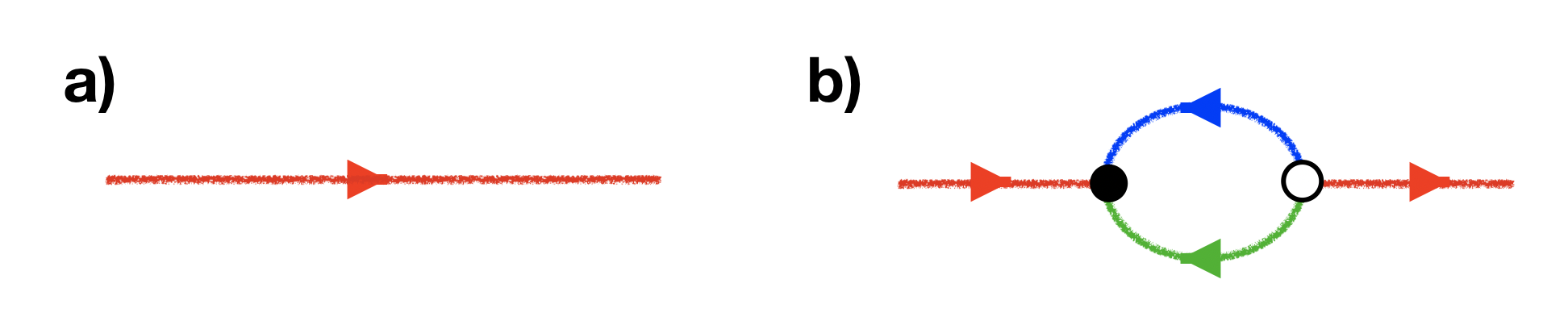}
\caption{\small Starting from e.g. a red line, we can nucleate a bubble described by a loop of green and blue edges.}
\label{fig:bubbling}
\end{center}
\end{figure}

In order to continue the study of abstract graphs, we now focus on relations between different graphs, and the possible existence of local operations in a graph which generate more complicated ones. We focus on the simplest kind, which we refer to as bubbling, and whose simplest version already appeared in the previous sections, see Figure \ref{fig:bubbling}. Let us note that such operations change the topology of the graph, but are in general physically allowed as discussed in section \ref{sec:stability}.

\begin{figure}[htb]
\begin{center}
\includegraphics[scale=.4]{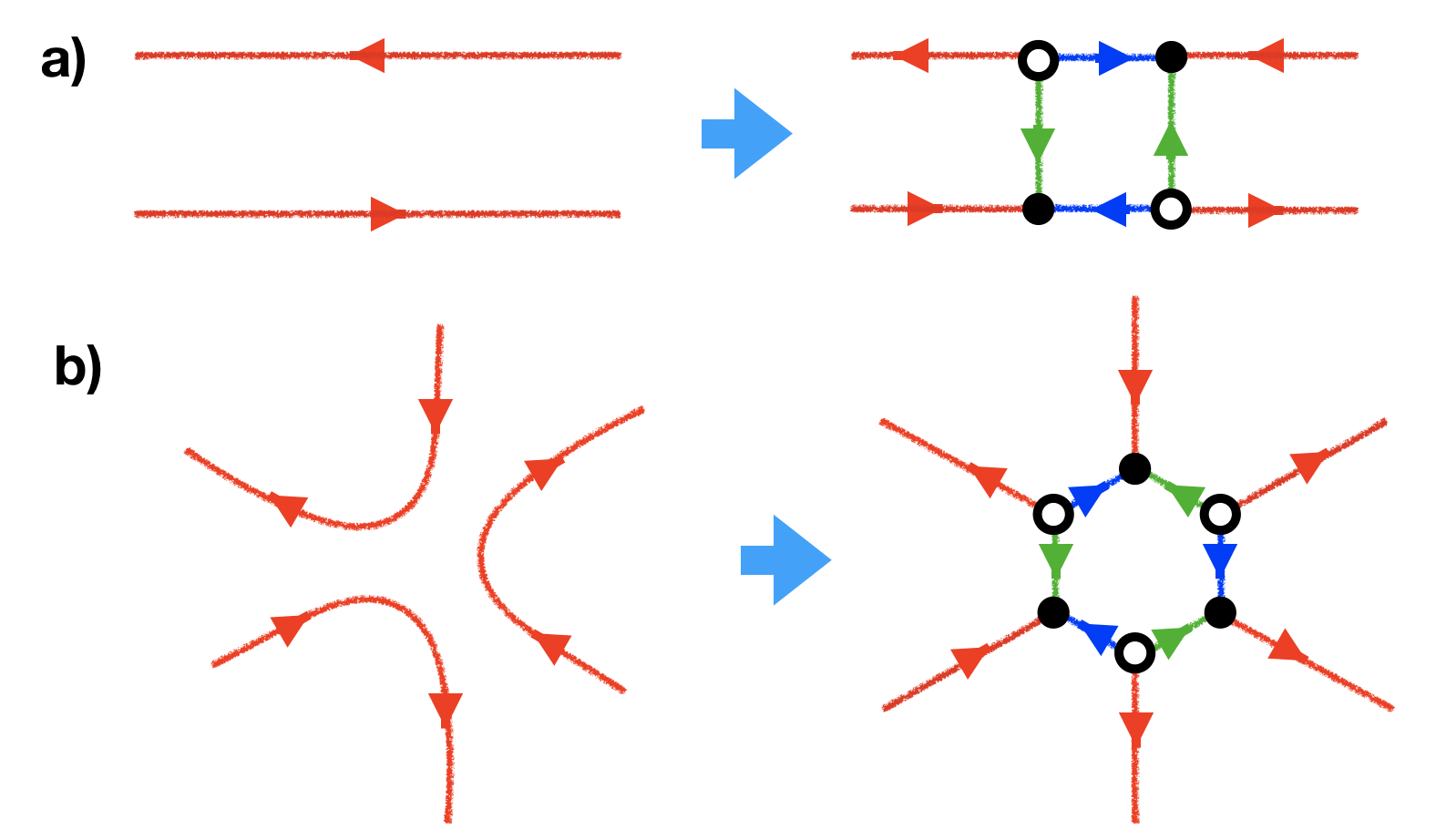}
\caption{\small Starting from $m$ red lines, we can nucleate a bubble described by a loop of a $(2m)$-sided polygon with alternating green and blue edges. Here we show the examples of $m=2$ (figure a) and $m=3$ (figure b).}
\label{fig:bubbling-more}
\end{center}
\end{figure}

Actually this is the simplest example of starting with $m$ red lines and bubbling a polygon with $2m$ sides with vertices attached to the (now) $2m$ red semi-infinite edges, and with $m$  internal alternating blue and green edges, see some examples in Figure \ref{fig:bubbling-more}.

\begin{figure}[htb]
\begin{center}
\includegraphics[scale=.35]{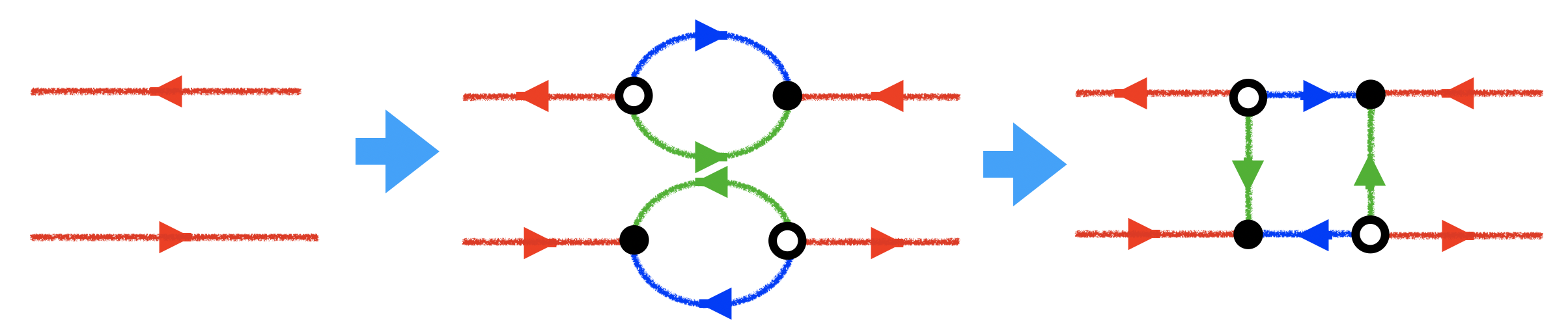}
\caption{\small Generating  the square in Figure \ref{fig:bubbling-more}a from the simple bubbling of Figure \ref{fig:bubbling}. In the first arrow we use twice the simple bubbling of Figure \ref{fig:bubbling}, and in the second arrow we recombine the green edges.}
\label{fig:generating-bubbles}
\end{center}
\end{figure}

The underlying idea is that there is a minimal set of operations which, starting with a trivial graph, generates all possible ones. For instance, for the above sets of bubbling, we can always generate them with the basic bubbling of Figure \ref{fig:bubbling} and a basic recombination of edges of same color and opposite orientation. This is illustrated in Figure \ref{fig:generating-bubbles}, where we show how to generate the square in Figure \ref{fig:bubbling-more}a from the simple bubbling of Figure \ref{fig:bubbling}. 

\begin{figure}[htb]
\begin{center}
\includegraphics[scale=.45]{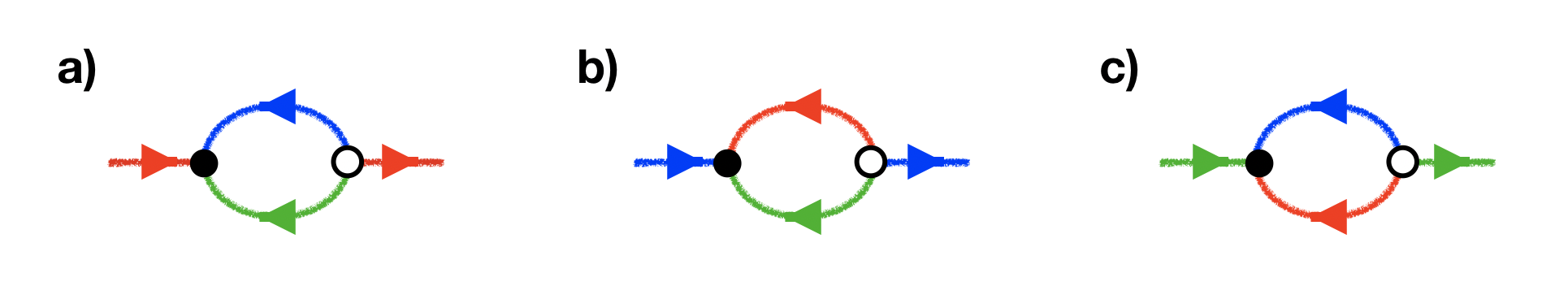}
\caption{\small The three basic bubblings. They are obtained from that in Figure \ref{fig:bubbling} by the color flip of suitable paths, recall Figure \ref{fig:path-color-flip1}.}
\label{fig:basic-bubblings}
\end{center}
\end{figure}

Of course, the bubbling of red lines is not the only possible bubbling process. We also have the color flipped versions of that, namely changing the external legs to blue or to green, and changing one of the internal edges accordingly, recall  Figure \ref{fig:path-color-flip1}. These three basic bubbling diagrams are shown in Figure \ref{fig:basic-bubblings}. It is clear that these basic bubbling diagrams, together with simple recombinations, suffice to generate all possible graphs of the theory. 

With these intuitions about the rich structure of the general set of graphs, we turn to their topological classification.

\subsection{Topological classification of graphs}
\label{sec:graph-classification}

The idea is to provide a completely general description of the topological structure of the heterotic networks. We will use the graph theoretical language of the previous section.
 We are going to focus on compact graphs, namely with no semi-infinite edges. We skip the fairly straightforward generalization of non-compact graphs, and simply use it in some concrete explicit examples.

\subsubsection{The red sub-graph}
\label{sec:red}

We aim to describe the structure of the graph in a way ultimately adapted to a generalized version of the construction in section \ref{sec:review}. In this description, the graph is obtained after a $\IZ_2$ quotient of a parent space graph, and the behavior of red edges is different from those of blue and green edges. In particular, each red edge arises from a red circle, which is turned into an interval by the action of the $\IZ_2$, with the two fixed points becoming the vertices.

Let us focus on the red edges and proceed as follows. Take the initial graph and erase all the green and blue edges, so we are left with only the set of red edges, and the vertices attached to them. Since it is a perfect matching, all the vertices of the graph are kept. We obtain a set of disconnected segments, which we label as $a=1,\ldots, E_R$. In the $\IZ_2$ covering space, the graph is topologically a finite disconnected set of $E_R$ red circles, with two marked points (the ones becoming white or black nodes). We now label them with $a=1,\ldots,2E_R$. For convenience we order them in a line, according to this label, and call this an ordered representation of the red edges. This is very convenient to regard the $\IZ_2$ as acting as a reflection with respect to this line, turning all the red circles into segments, and the fixed points into the location of the vertices. 

\begin{figure}[htb]
\begin{center}
\includegraphics[scale=.5]{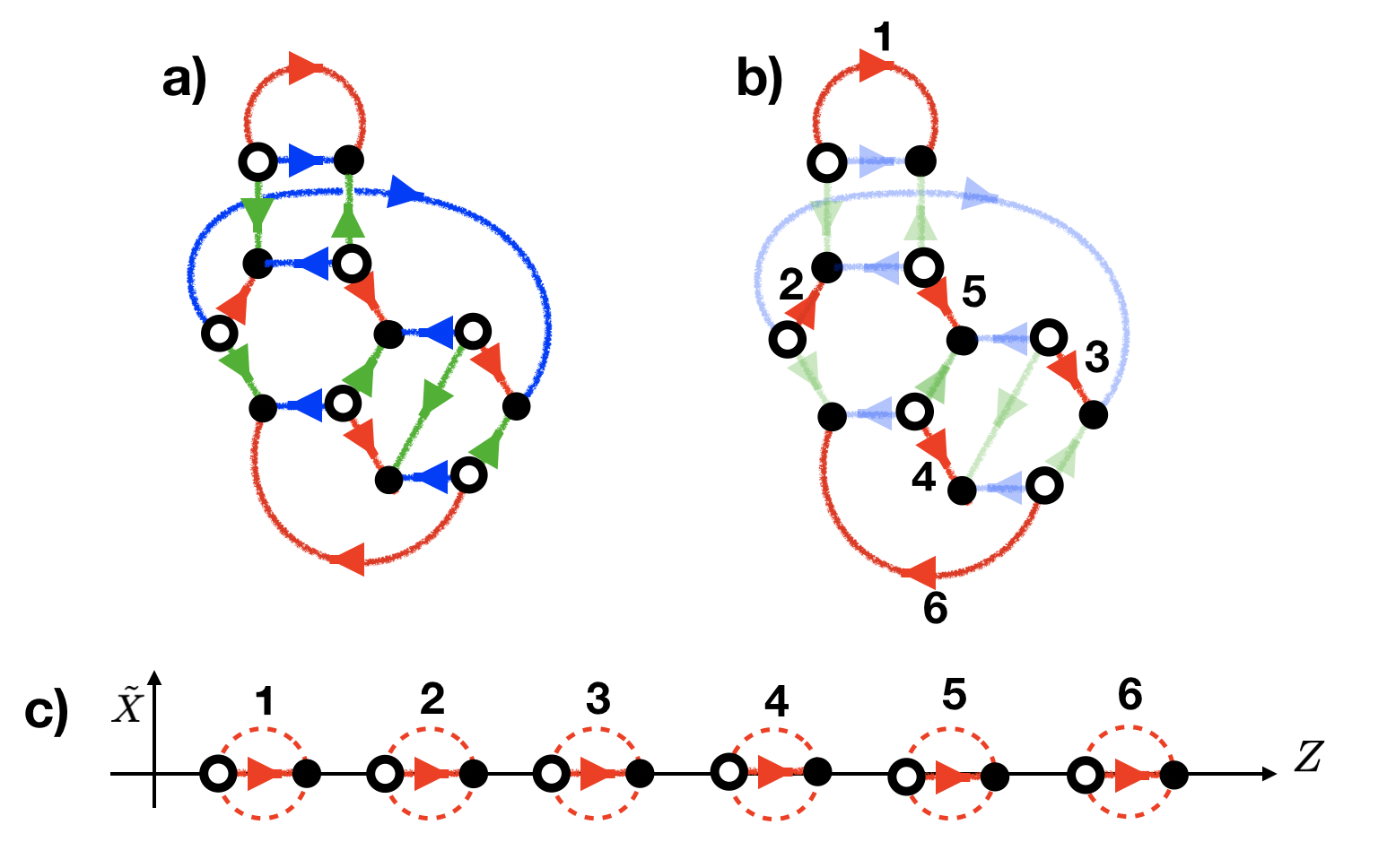}
\caption{\small Structure of the red subgraph for an example of a pretty general compact graph. We start with Figure (a) and remove all green and blue edges, leaving only the red ones, which we label $1,\ldots,E_R$, with $E_R=6$ in this example, Figure (b). Then in Figure (c) we show the ordered representation of the edges, for the chosen labeling: we align the red edges in the real line, according to their labels, and with white and left vertices at the left and right of the edge, respectively. The red dashed lines indicate the red circles which correspond to the edges when regarded in the $\IZ_2$ covering space. We have also indicated the $Z$ and ${\tilde X}$ axes, which will allow for an easy algebraic embedding of the whole set of red circles, in section \ref{sec:graph-algebraic}.}
\label{fig:red-subgraph}
\end{center}
\end{figure}

This is shown in Figure \ref{fig:red-subgraph}. Let us mention with hindsight that the configuration will admit an embedding as a set of circles in a 2-plane, with coordinates $(Z,{\tilde X})$, such that the $\IZ_2$ action is ${\tilde X}\to -{\tilde X}$. This means that the vertices are going to be located at ${\tilde X}=0$, namely the horizontal line, as suggested above. 

The point to emphasize is that, if we focus on just the red edges, the only relevant information about the graph is their number. They are just disconnected, with no information about the connectivity of the graph. This will be provided by the green and blue edges, which comes next.

\subsubsection{The green/blue subgraph}
\label{sec:greenblue}

In order to encode the information of green and blue edges, we proceed as follows. Take the initial graph and form the $G{\ov B}$ path, namely erase all the red edges and leave only the green and blue edges, and flip the orientation of the blue ones (for $B{\ov G}$ paths, we simply get the orientation reversed story). 

\begin{figure}[htb]
\begin{center}
\includegraphics[scale=.5]{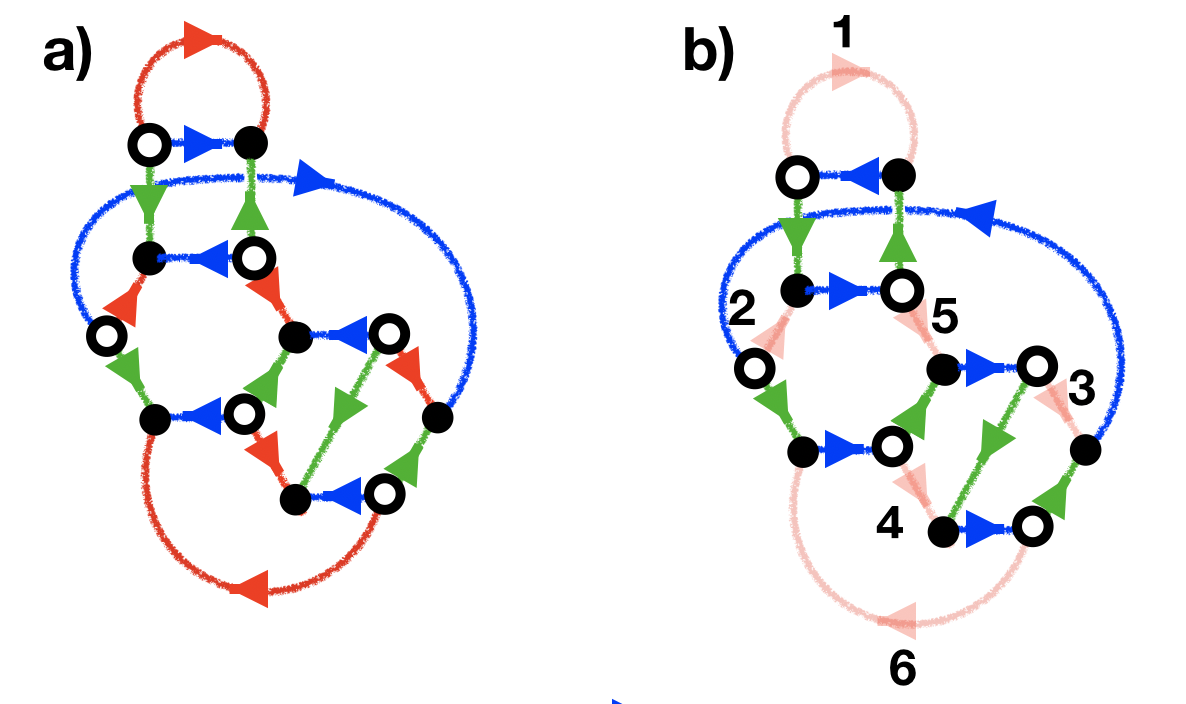}
\caption{\small Structure of the green/blue subgraph for the example of the graph in Figure \ref{fig:red-subgraph}. The full graphs is shown in (a) and its $G{\ov B}$ paths are shown in Figure (b). The latter displays two closed $B{\ov G}$ loops, one with 4 edges (intersecting the red edges in the ordering $(1,1,2,5)$), and another with 8 edges (intersecting the red edges in the ordering $(2,6,4,5,3,4,6,3)$.}
\label{fig:blue-green-subgraph}
\end{center}
\end{figure}

The $G{\ov B}$ path defines a set of closed loops (including the possibility of getting just one single connected component), which in total pass through all the vertices in the graph. Also, each green/blue connected component touches a subset of the red edges; hence, remembering the labeling of red edges it defines an ordered set of integers $(n_1,n_2,\ldots,n_{2p})$ (up to cyclic permutations), with each entry taking values in $\{1,\ldots, E_R\}$, where $2p$ is the total number of  edges in the path (half green, half blue). Note also that if a path touches a red edge in a node, there is no other path touching that node (because all of its valence has been occupied). Finally, note that if a path touches a red edge at a white node, it will touch the next one (in the ordering $(n_1,n_2,\ldots,n_{2p})$) at a black node. Finally, we recall that every time the path touches a vertex, it flips from blue to green and viceversa. The structure of the $G{\ov B}$ path and its different components is shown in Figure \ref{fig:blue-green-subgraph} for the same graph as in Figure \ref{fig:red-subgraph}.

The sets of ordered integers $(n_1,n_2,\ldots,n_{2p})$ for the different paths (which in general have different $p$'s) encodes all the information about the connectivity of the graph. We can then represent how these paths intersect the different red edges (or circles, in the $\IZ_2$ covering) in the ordered representation of the red edges. Each path corresponds to a closed curve which passes in an alternating fashion through white and black vertices of the red edges in the ordering $(n_1,n_2,\ldots,n_{2p})$ for the $p$ corresponding to that path. A useful analogy is that the path is like a thread, passing though ``holes'' (the vertices) from one side  of the fabric (the plane containing the red circles) to the other (i.e. through while of black nodes) and viceversa, with the ordering $(n_1,n_2,\ldots,n_{2p})$, and closing onto itself after $2p$ steps. The result is that the path `sews' together a subset of the red circles. Note that the paths are mutually exclusive: each hole must be used once and only once, so you cannot pass a thread through a hole if it has been already used by some other path.

\begin{figure}[htb]
\begin{center}
\includegraphics[scale=.45]{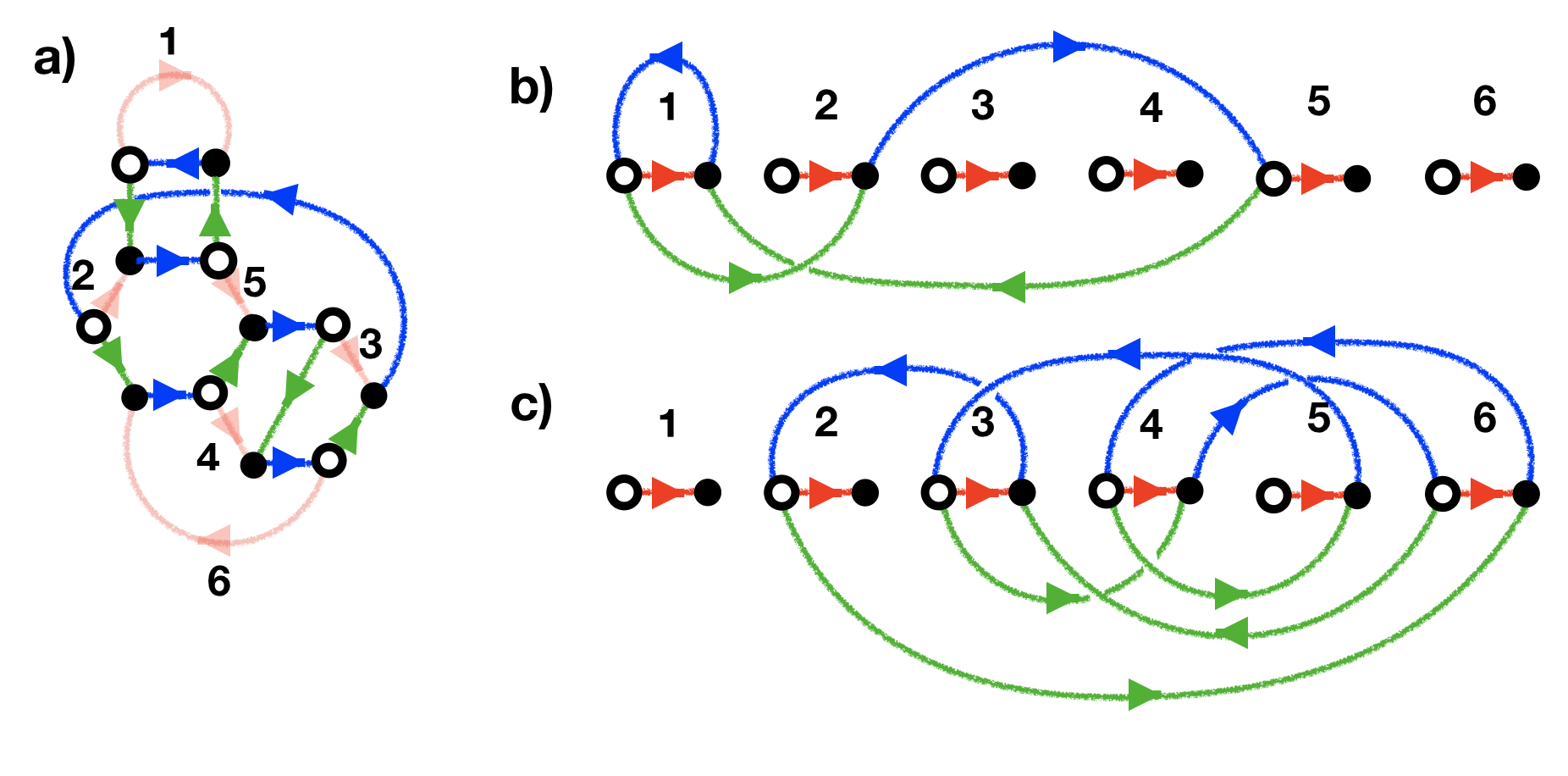}
\caption{\small Structure of the representation of ordered red edges and sewn green/blue paths for the example of the graph in Figure \ref{fig:red-subgraph}. The diagram (a) is just that in Figure \ref{fig:blue-green-subgraph}b, repeated for convenience. Figures (b) and (c) depict the green/blue loops, of 4 and 8 edges in this example. Note that they reproduce the intersections with red edges in the orders $(1,1,2,5)$ and $(2,6,4,5,3,4,6,3)$, respectively, and alternating between white and black nodes.}
\label{fig:threads}
\end{center}
\end{figure}

We refer to the structure of the paths and how the interplay with the ordered representation of the red edges as `ordered \& sewn' representation. In Figure \ref{fig:threads} we show it for the graph in Figure  \ref{fig:blue-green-subgraph}.

The representation above reproduces the connectivity correctly. It should be possible to quantify more precisely the combinatorics of the green/blue paths and extract e.g. an estimate of the number of possible graphs, perhaps even refined according to the number of connected components of the green/blue paths, and their lengths. We will not attempt to do so in the present work, and continue elaborating on the features arising for a generic graph.

\begin{figure}[htb]
\begin{center}
\includegraphics[scale=.45]{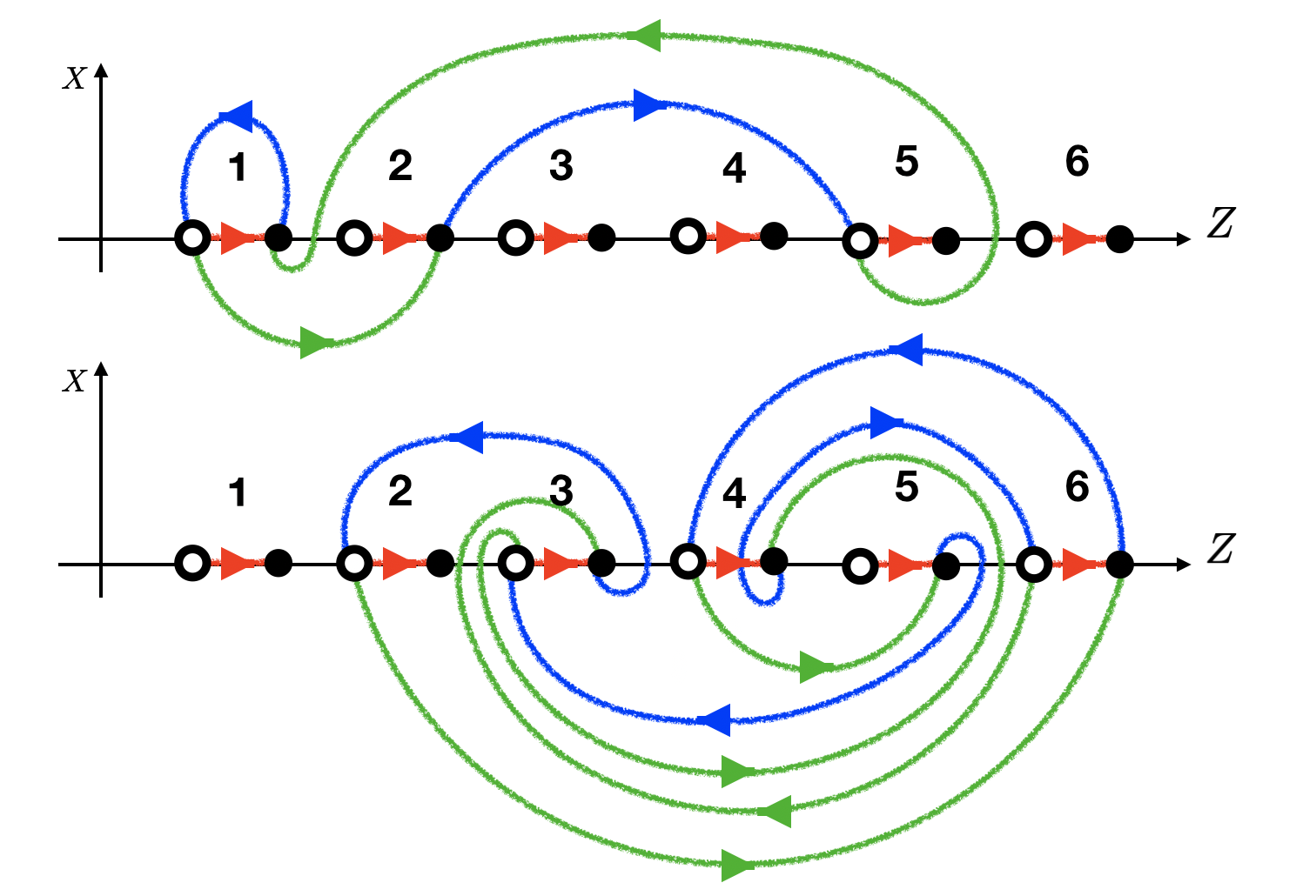}
\caption{\small A representation of the green/blue paths of Figure \ref{fig:blue-green-subgraph}b, c in terms of closed curves without self-intersections. With hindsight for the forthcoming algebraic embedding, we have depicted the $Z$ and $X$ axes to emphasize that these curves are located at ${\tilde X}=0$, which corresponds to the plane of the picture. The ${\tilde X}$-plane should be regarded as orthogonal to the picture, towards the viewer; then, the red edges can be regarded as red circles in the $(Z,{\tilde X})$-plane, viewed from `above'.}
\label{fig:jordan}
\end{center}
\end{figure}

One minor drawback when one attempts to represent the above green/blue curves in a 2-plane (like the $(Z,X)$-plane used later) is that in general they seem to have self-intersections when represented on the plane. Even if they would be present, self-intersections is a minor drawback because they actually correspond to a non-chiral junction, which we know can be resolved (this perhaps needs to be done by adding one extra dimension). But actually, the apparent self-intersections are just an artifact of the simplified picture with blue and green lines on different sides of the horizontal line.  It is actually possible in general to twist the green/blue curve into a sufficiently curled up `snake' such that there are no self-intersections. This is shown for the above example in Figure \ref{fig:jordan}, where we have depicted the axis $Z$ and $X$, with hindsight for the forthcoming algebraic embedding.

We now turn to the algebraic description of the embedding of this graph into an ambient space, and its realization as a $(0,1)$ worldsheet theory. 

\subsection{Algebraic realization and worldsheet description}
\label{sec:graph-algebraic}

We now provide a description of the graph as the zero locus of a set of equations. The problem of embedding a graph in terms of algebraic equations is a mathematically well studied question in real algebraic geometry, with powerful theorems (see \cite{Akbulut-King, Akbulut} for reviews) which guarantee that such embedding is always possible in a sufficiently large number of dimensions. One important point is that there exist many possible ways to achieve such an embedding.
In this section we will focus on a particularly simple one, which is well suited for the realization of the network in terms of a 2d $(0,1)$ worldsheet theory.

The procedure follows the earlier construction in section \ref{sec:graph-classification}, embedding it into a generalization of the construction in section \ref{sec:review}. In particular, we consider two real coordinates $X$ and ${\tilde X}$, with ${\tilde X}$ being odd under a $\IZ_2$ gauge symmetry, along with a coordinate $Z$, and the additional 9d coordinates which will be mere spectators. We consider the red sub-graph introduced in section \ref{sec:red}, regarded in the $\IZ_2$ covering space, and we embed it in the $(Z,{\tilde X})$-plane via the expression
\beqa
{\tilde X}^2+f(Z)=0\, ,
\label{red-circles}
\eeqa
where $f(Z$) is a polynomial with a number of zeroes given by $2E_R$, twice the number of red edges  in the graph (i.e. of red circles in the $\IZ_2$ covering space). A simple choice is to let the zeroes be located at the points $(Z,{\tilde X})=(n,0)$ with $n=1,2,\ldots, 2E_R$. A minimal choice is
namely
\beqa
f(Z)=(Z-1)(Z-2)\ldots (Z-2E_R)\, ,
\eeqa
where the overall sign is fixed so that the $f(Z)$ has negative slope at the zero $Z=1$ (or any odd value), and positive slope at $Z=2$ (or any even value). Namely ${\rm sign}(f'(Z=n))=(-1)^{n}$. For instance, locally near $Z=1$, we have $f(Z)\sim -\delta Z\equiv -(Z-1)$, and (\ref{red-circles}) becomes $\delta Z\sim {\tilde X}^2$, i.e. is a parabola open towards positive $Z$, while locally near $Z=2$ we have $g(Z)\sim +\delta Z\equiv Z-2$ and  (\ref{red-circles}) becomes $\delta Z\sim -{\tilde X}^2$, i.e. a parabola open towards negative $Z$. The picture gets very clear by taking a quadratic $f(Z)\sim (Z-1)(Z-2)$ (with zeroes at $Z=1,2$), which reproduces the equation of a (round circle) of diameter 1 centered at $Z=3/2$. The general picture is very similar to the discussion in section \ref{sec:more-bubbles}.

In general, we may be interested in non-compact graphs, so that some of the red circles actually reach out to infinity. This can be easily accomplished by minor modifications of the above picture. To include all those cases, we will describe the red curves as the zero locus of a function in the $({\tilde X},Z)$-plane, namely
\beqa
F(Z,{\tilde X})=0\, ,
\label{red-curves}
\eeqa
with the proviso that the function is even under ${\tilde X}\to -{\tilde X}$.

We now describe the green/blue curve in the $(Z,X)$-plane. As we have seen, this can have a very non-trivial form, and in general it is not feasible to provide an explicit expression for it (see later for some concrete examples). However, each independent component is a closed continuous curve with no self-intersections, namely a Jordan curve (i.e. an injective map $\IS^1\to \IR^2$), and general results guarantee that for sufficiently smooth curves, they can be expressed as the zero of a continuous function defined over $\IR^2$. So we describe the green/blue curve by
\beqa
G(Z,X)=0\, ,
\label{bluegreen-curves}
\eeqa
(where this may be a product of expressions describing the different connected components). In addition we have the selector constraint 
\beqa
X{\tilde X}=0\, ,
\label{selector}
\eeqa
which selects the curves to be embedded in the $(Z,X)$- or the $(Z,{\tilde X})$-planes. This is very close to a generalization of the construction in section \ref{sec:review}, save for the fact that need to combine the conditions (\ref{red-curves}) and (\ref{bluegreen-curves}) into a single one. In general there is no unique way to do this, but a simple approach is to look for a function $H(Z,{\tilde X},X)$ which satisfies
\beqa
H(Z,{\tilde X},X=0)=F(Z,{\tilde X})\quad , \quad H(Z,{\tilde X}=0,X)=G(Z,X)\, .
\label{extension-conditions}
\eeqa
This is possible if both functions are compatible, namely 
\beqa
F(Z,{\tilde X}=0)=G(Z,X=0)\, ,
\label{compatibility}
\eeqa
(because both must be equal to $H(Z,{\tilde X}=0,X=0)$). A simple function satisfying this is
\beqa
H(Z,{\tilde X},X)=F(Z,{\tilde X})+G(Z,X)-F(Z,{\tilde X}=0)\, ,
\label{def-h}
\eeqa
which clearly satisfies (\ref{extension-conditions}).

We are now ready to construct the worldsheet description of the network of heterotic theories. As in section \ref{sec:review}, we introduce three right-moving multiplets $Z,X,{\tilde X}$, in addition to the 9d ones, and (in order to cancel 2d gravitational anomalies) two Fermi multiplets $\Lambda,{\tilde\Lambda}$. The tilded fields are odd under a gauge $\IZ_2$ (which is anomaly free because it acts non-chirally on the set of left- and right-moving fermions). Finally we have a superpotential
\beqa
W=\Lambda H(Z,{\tilde X}, X)+{\tilde \Lambda}X{\tilde X}\, .
\label{supo-h}
\eeqa
Hence, the vacuum equations require (\ref{selector}) and
\beqa
H(Z,{\tilde X},X)=0\, ,
\eeqa
so that for $X=0$, ${\tilde X}\neq 0$ we obtain (\ref{red-curves}), i.e. the red curves, while for ${\tilde X}=0$, $X\neq 0$ we obtain (\ref{bluegreen-curves}), i.e. the green/blue curves. Note the interesting feature that the compatibility condition (\ref{compatibility}) geometrically means that the endpoints of the red edges (which are the points with $X=0$, ${\tilde X}=0$ in the red circles in the $\IZ_2$ covering space, namely satisfying $X=0$, $F(Z,{\tilde X}=0)=0$) are points which also belong to the green/blue curve (namely, satisfy ${\tilde X}=0$, $G(Z,X=0)=0$).

We can furthermore show that the worldsheet theory near one of these junction points is exactly of the local form of a basic trivalent junction. Expanding around a point $X={\tilde X}=0$ satisfying $H(Z_0,{\tilde X}=0,X=0)=0$ we have
\beqa
H(Z,{\tilde X},X)\sim (Z-Z_0) + \frac{\partial^2 H}{\partial {\tilde X}^2}\Bigg|_{{\tilde X}=0}\, {\tilde X}^2+ \frac{\partial H}{\partial X}\Bigg|_{ X=0}\, X\, ,
\eeqa
where we have used that the symmetry ${\tilde X}\to -{\tilde X}$ prevents linear terms in ${\tilde X}$ (but not in $X$). This reproduces a junction of the form considered at the end of section \ref{sec:conjugate}, with linear dependence in $X$. In case the linear term in $X$ happens to vanish, the quadratic term in the Taylor expansion lead to the same junction in the form in \cite{Tachikawa:2026top}, reviewed in section \ref{sec:review}.

\subsection{Some examples}
\label{sec:graph-examples}

In this section we work out some examples of the above procedure. For instance, let us recover the description of the basic junction of section \ref{sec:review}, c.f. Figure \ref{fig:heterotic-junction}b. This is obtained by a straightforward extension of the above recipe to include non-compact edges. The red semi-infinite edge can be described as a parabola in the $(Z,{\tilde X})$-plane, open towards $Z\to \infty$, namely
\beqa
F(Z,{\tilde X})={\tilde X}^2-Z=0\, ,
\label{junction-f}
\eeqa 
in the notation of the previous section. On the other hand, the green/blue curve can be written as a parabola in the $(Z,X)$-plane, open towards $Z\to -\infty$, namely
\beqa
G(Z,X)=-X^2-Z=0\, .
\label{junction-g}
\eeqa
Both functions are compatible, namely $F(Z,{\tilde X}=0)=G(Z,X=0)$, so both can be combined into (\ref{def-h})
\beqa
H(Z,{\tilde X},X)={\tilde X}^2-X^2-Z\, ,
\label{junction-h}
\eeqa
and the superpotential (\ref{supo-h}) reproduces (\ref{tachikawa-supo}), and hence the construction in \cite{Tachikawa:2026top}. As explained in section \ref{sec:conjugate}, the conjugate junction is simply obtained by flipping the sign of $Z$, in our case in  (\ref{junction-f}) and (\ref{junction-g}), and hence (because they are compatible) in (\ref{junction-h}). 

It is also straightforward to construct the bubbling diagrams of sections \ref{sec:gluing}, \ref{sec:more-bubbles}, by simply taking
\beqa
F(Z,{\tilde X})={\tilde X}^2-f(Z)\quad ,\quad G(Z,X)=-X^2-f(Z)\, ,
\eeqa
with $f(Z)$ a function with $n$ zeroes. 

Finally, the compact networks of section \ref{sec:compact} are easily obtained by similar expressions, by simply flipping the sign of the $X^2$ term. It is straightforward to work out other examples of networks, but we refrain from doing so and move on to develop new efficient ways to describe fairly complicated networks with minor modification of the above techniques.

\section{Adding extra sheets}
\label{sec:sheets}

The construction in the previous section provides an algebraic embedding of a general graph in an ambient space amenable for the construction of a $(0,1)$ heterotic worldsheet realizing the network. The procedure can however quickly become fairly complicated in practice, as the green/blue curves can become highly non-trivial. On the other hand, as already emphasized, in real algebraic geometry the realization of a graph as the zero set of some algebraic functions is far from unique. This suggests the possibility of finding a simpler algebraic realization of some classes of graphs, or even more in general, of devising some optimization algorithm to find the simplest possible realizations for each graph.

We do not have a complete answer to this general problem. However, in this section we show that a particular class of networks admits a more efficient description than that provided by the general algorithm of the previous section. We will introduce this class of networks via some illustrative examples, and subsequently proceed to their general definition in graph-theoretical terms.

\subsection{The $X$-sheets}
\label{sec:x-sheets}

The key idea is that it is possible to modify the selector condition $X{\tilde X}=0$ to a more general expression
\beqa
f_n(X){\tilde X}=0\, ,
\label{fn-x-sheets}
\eeqa
where $f_n(X)$ is a (e.g. polynomial) function with $n$ zeroes $X_i$. This implies that the branch with ${\tilde X}=0$ is actually realized in $n$ different copies, located at $X=X_i$, which we dub the $X$-sheets. Let us illustrate this with some examples.

\subsubsection{The multiple-sheeted junction}

Consider
\beqa
W=\Lambda ({\tilde X}^2-X^2-Z) + {\tilde\Lambda}(X^2-X_0^2){\tilde X}\, .
\eeqa
The vacuum equations are
\beqa
&(X^2-X_0^2){\tilde X}=0 \, ,&\nonumber\\
& {\tilde X}^2-X^2=Z\, , &
\eeqa
namely of the kind considered above, for the case of 2 $X$-sheets. The structure of branches has several components:

$\bullet$ As explained, the solutions with ${\tilde X}\neq 0$ are a set with two components $X=\pm X_0$, each one described by ${\tilde X}=\pm \sqrt{Z+X_0^2}$. These two real curves, which we call ${\tilde C}_\pm$, exist for $Z>-X_0^2$ and are disjoint red parabolas semi-infinite towards large  positive $Z$ and large $|{\tilde X}|$. 

$\bullet$ We also have solutions with ${\tilde X}=0$, and then $X=\pm \sqrt{-Z}$. This curve, which we call $C$, exists for $Z<0$, describes a parabola semi-infinite towards large negative $Z$ and large $|X|$.

\begin{figure}[htb]
\begin{center}
\includegraphics[scale=.4]{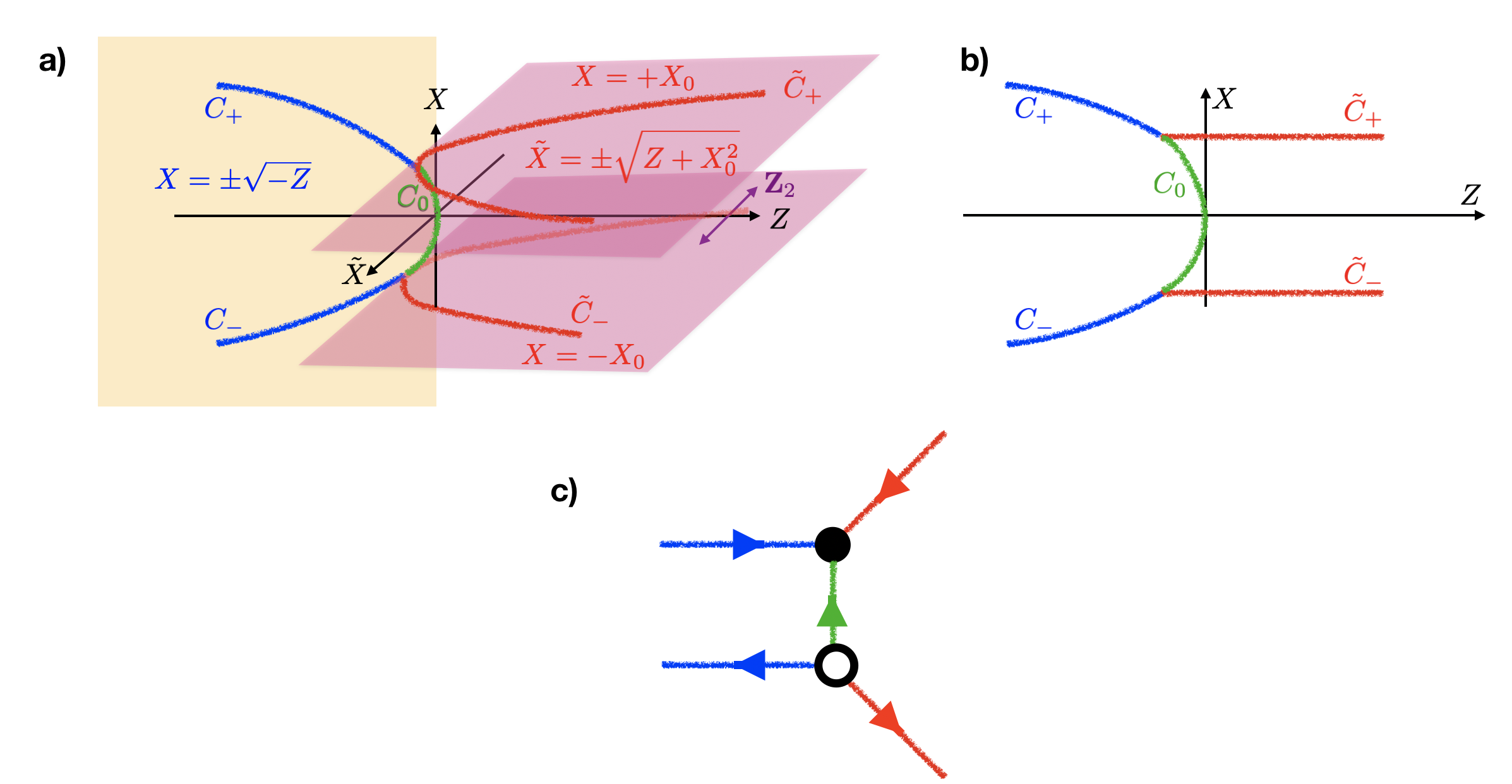}
\caption{\small a) The network with the two branches ${\tilde C}_\pm$ (in red), and the green/blue curve split into a middle piece $C_0$ (in green) and two semi-infinite pieces $C_\pm$ (in blue), in the $\IZ_2$ covering space b) A simplified depiction after the $\IZ_2$ quotient. c) Graph depiction of this network.}
\label{fig:two-two}
\end{center}
\end{figure}

The curve $C$ intersects the curves ${\tilde C}_\pm$ at $X=\pm X_0$, ${\tilde X}=0$, $Z=-X_0^2$. The configuration therefore has 2 junction points, and the curve $C$ is split into three pieces, as depicted in Figure \ref{fig:two-two}, with a finite piece which we denote by $C_0$ and two semi-infinite pieces denoted by $C_\pm$.

One may be surprised that the asymptotic branches in $C_\pm$ carry the same theory, as they look analogous to those of the basic junction, in which they have different colors. In fact, it is easy to compare the masses of fermions which are integrated out in them and check that there is no relative $q$ factor that arises. In other words, there is no flip of the sign of the fermion masses. To see this more explicitly, in the ${\tilde X}=0$ branch the fermions masses read
\beqa
\lambda\left(-2X\psi_X-\psi_Z\right)
+\widetilde{\lambda}\left(X^2-X_0^2\right)\psi_{\widetilde{X}}\, .
\eeqa
This expression makes manifest the change of sign of the second term between the regions
\beqa
X\in(-\infty,-X_0)\cup(X_0,+\infty)\, ,
\eeqa
and
\beqa
X\in(-X_0,X_0)\, ,
\eeqa
assuming $X_0>0$. We see that, even though the branches $C_\pm$ differ in the sign of $X$, this does not change the mass of the fermions, which is controlled by $X^2$. 

\subsubsection{Polygon networks}

It is now easy to provide more involved examples by combining these multiple-sheeted graphs, in analogy with the bubbles in section \ref{sec:gluing}, \ref{sec:more-bubbles}, by promoting $Z$ to a more general function. For instance, consider 
\beqa
W=\Lambda ({\tilde X}^2-X^2-(Z^2-a)) + {\tilde\Lambda}(X^2-X_0^2){\tilde X}\, .
\eeqa
The vacuum equations are
\beqa
&(X^2-X_0^2){\tilde X}=0 \, ,&\nonumber\\
& {\tilde X}^2-X^2=Z^2-a \, .&
\label{square}
\eeqa
We have the two branches $X=\pm X_0$, which in the $({\tilde X},Z)$-plane correspond to the hyperbola 
\beqa
{\tilde X}^2-Z^2=X_0^2-a\, ,
\eeqa
with the gap in one direction or the other depending on the sign of $X_0^2-a$.

\begin{figure}[htb]
\begin{center}
\includegraphics[scale=.5]{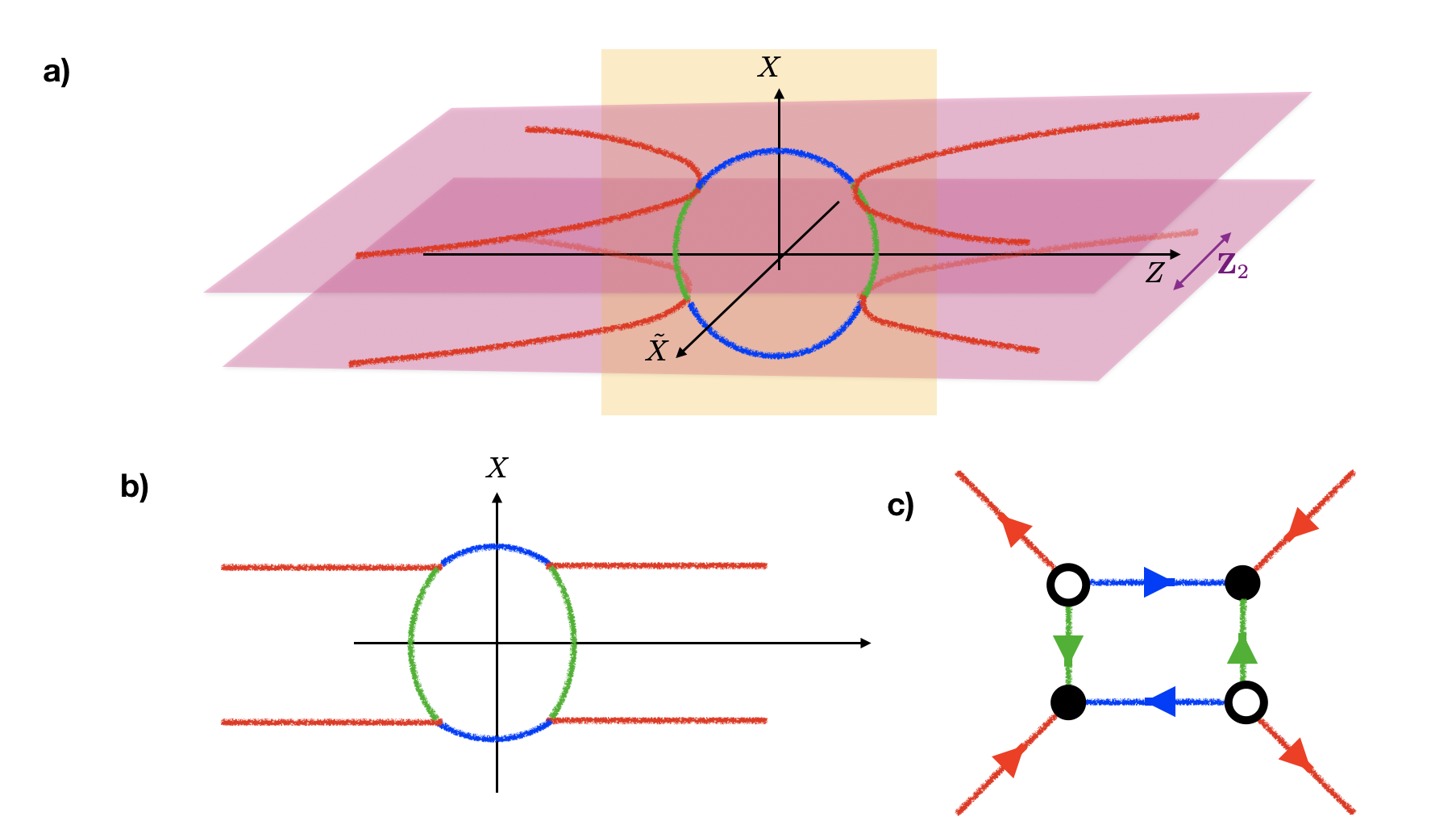}
\caption{\small a) The configuration of curves describing the square loop network. b) A schematic representation in the $\IZ_2$ quotient. c) The graph.}
\label{fig:square-loop-network}
\end{center}
\end{figure}

The other branch is ${\tilde X}=0$, which in the $(X,Z)$-plane is the circle
\beqa
X^2+Z^2=a\, .
\eeqa
The two branches touch at ${\tilde X}=0$, $X=\pm X_0$, and $Z=\pm\sqrt{a-X_0^2}$. Clearly we are interested in having intersections, so we focus on $a>X_0^2$. Then the geometry of the curves looks like Figure \ref{fig:square-loop-network}.

Clearly, using (\ref{fn-x-sheets}) with a polynomial with $n$ roots, we can construct an $n$-sheeted generalization of the above diagram in which the green/blue line is a polygon of $2n$ edges, and $2n$ junctions at which they join with $2n$ semi-infinite red lines.

\subsubsection{Ladder networks}

An alternative possibility is to flip the sign of the parabola in $Z$. This has the effect of conjugating the junctions, so that the red line (which now comes in $n$ copies) will correspond to the circle in the loop. Let us check this. Consider
\beqa
&f_n(X){\tilde X}=0 \, ,&\nonumber \\
&  {\tilde X}^2-X^2=-Z^2-a \, ,&
\eeqa
For concreteness we focus on $0<a <X_i^2$ for all $i$, with $X_i$ being the zeros of $f_n(X)$.

The branches $X=X_i$ lead to curves
\beqa
{\tilde X}^2+Z^2=-a+X_i^2\, .
\eeqa
This gives $n$ circles in the $({\tilde X},Z)$-plane, of different radii and located in different $X$-sheets.

\begin{figure}[htb]
\begin{center}
\includegraphics[scale=.5]{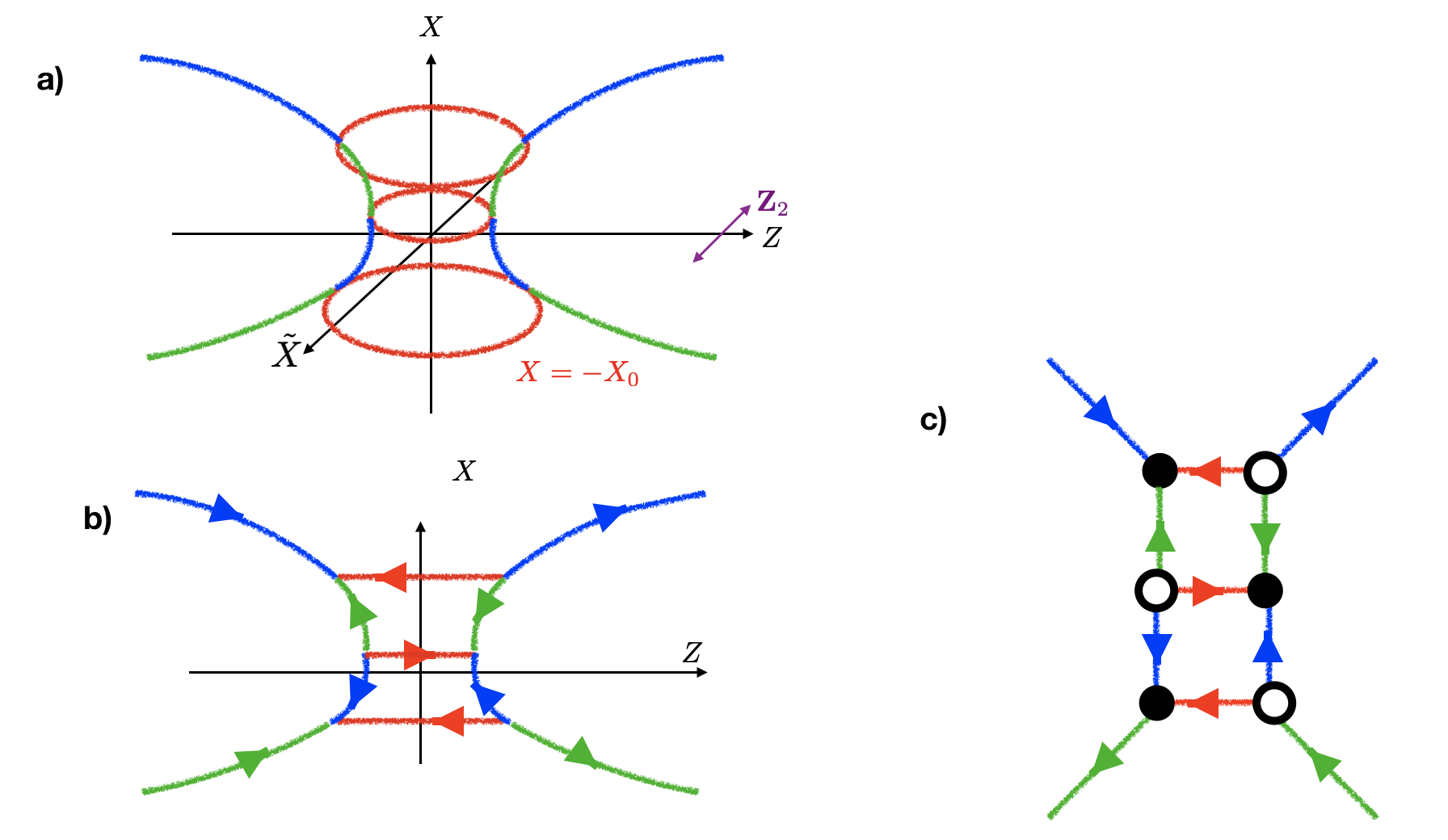}
\caption{\small a) Geometry of curves leading to a ladder network. b) A schematic representation in the $\IZ_2$ quotient. c) The graph}
\label{fig:ladder-network}
\end{center}
\end{figure}

The branch ${\tilde X}=0$ is given by a hyperbola in the $(X,Z)$-plane
\beqa
X^2-Z^2=a\, .&
\eeqa
Since $a>0$, this exists only for $|Z|^2>a$. 
This hyperbola intersects the circles at the points ${\tilde X}=0$, $X=X_i$, $Z=\pm\sqrt{-a+X_i^2}$. The resulting geometry and network (which we will call `ladder') are depicted in Figure \ref{fig:ladder-network}.

\begin{figure}[htb]
\begin{center}
\includegraphics[scale=.5]{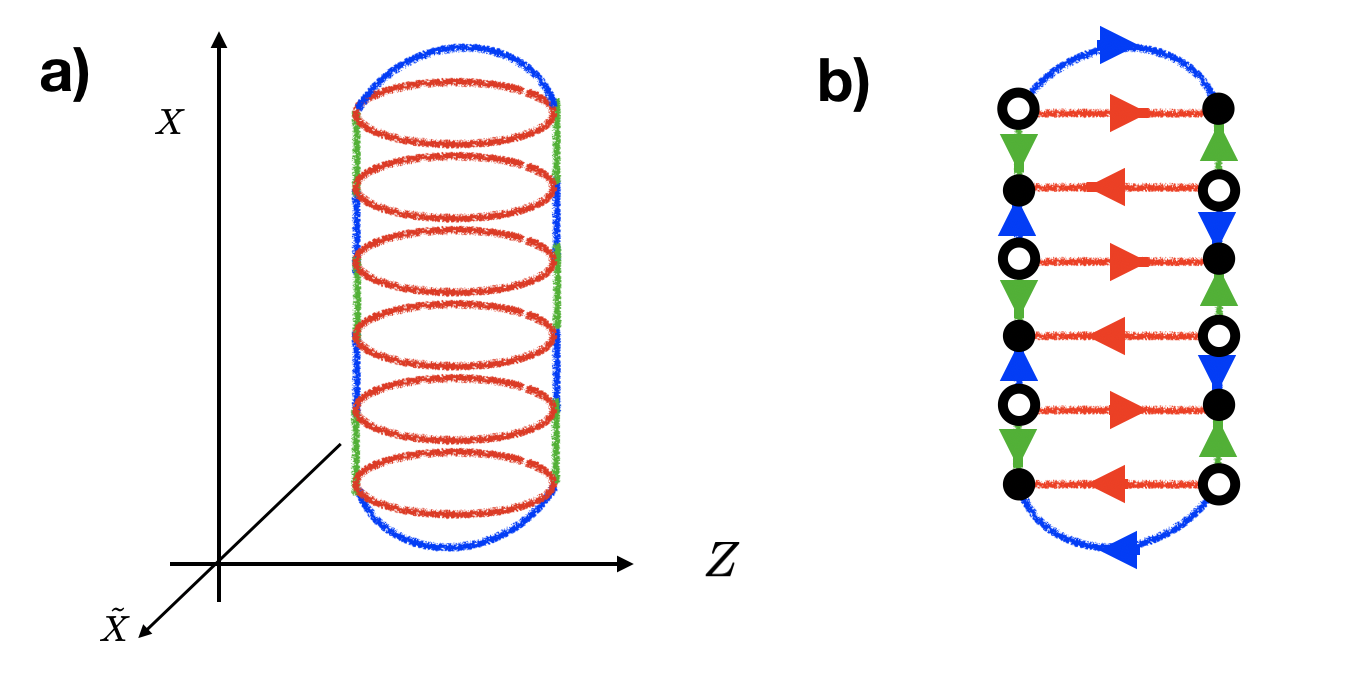}
\caption{\small a) Geometry of curves leading to a compact ladder network. b) The graph.}
\label{fig:compact-ladder}
\end{center}
\end{figure}

It is straightforward to construct compact examples of ladder networks, by simply flipping the sign of the $X^2$ term to remove the non-compact directions. We skip the hopefully by now familiar equations, and simply provide an illustrative example in Figure \ref{fig:compact-ladder}.

\subsubsection{Ladders and polygons}

\begin{figure}[htb]
\begin{center}
\includegraphics[scale=.5]{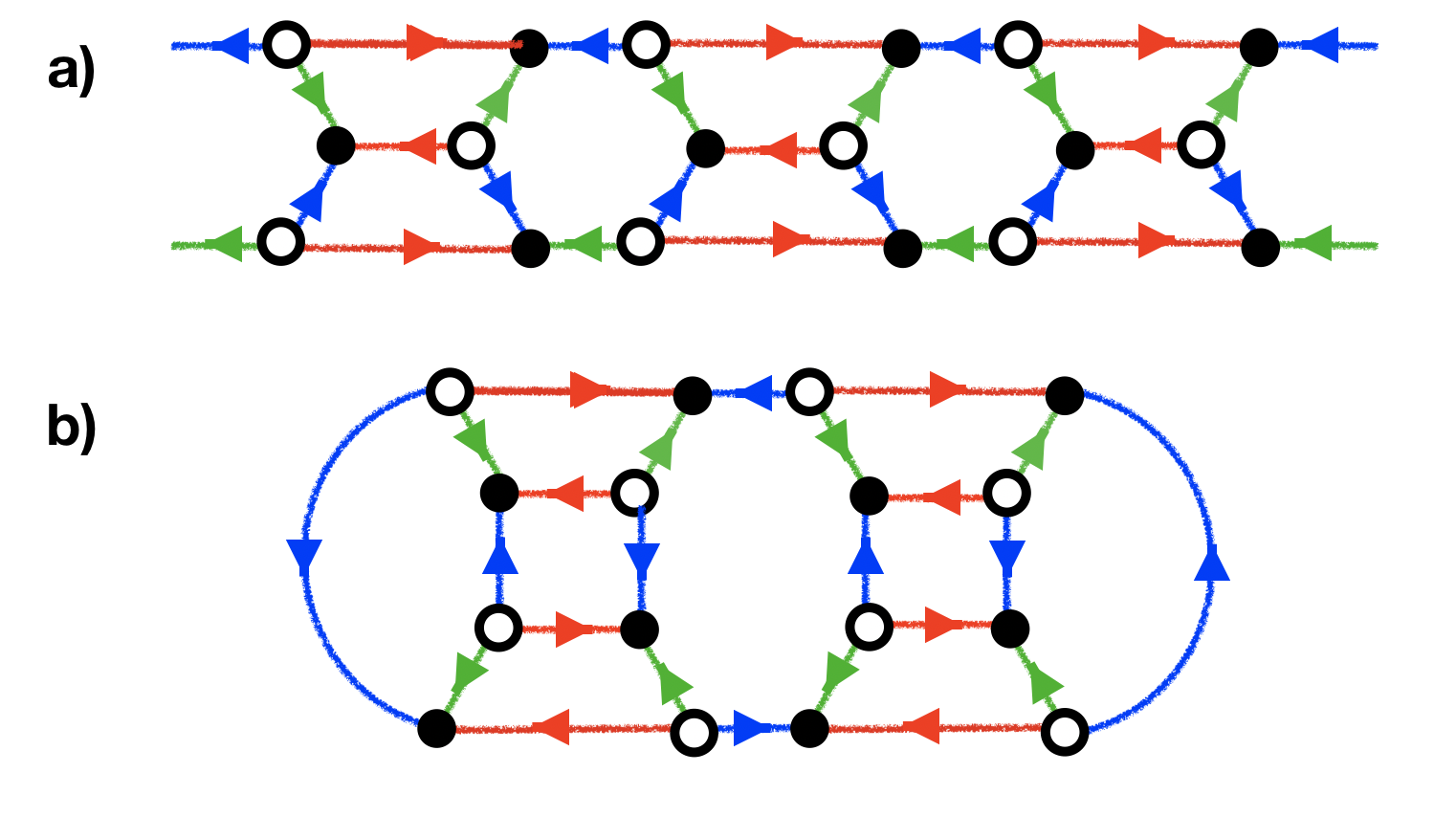}
\caption{\small a) Graph for a non-compact network combining ladders and polygons into a chain. b) Graph for a compact example.}
\label{fig:ladder-polygon}
\end{center}
\end{figure}

Finally, it is easy to obtain networks which combine arbitrarily long chains of ladders of $n$ red edges and $2n$-sided polygons of green/blue edges, by using the equations
\beqa
f_n(X){\tilde X}=0\quad ;\quad {\tilde X}^2\pm X^2-g_m(Z)=0\, ,
\eeqa
where the $\pm$ choice corresponds to compact/non-compact diagram, $f_n(X)$ is a function with $n$ zeroes, which introduces $n$ $X$-sheets, and $g_m(Z)$ is a function with $m$ zeroes, each describing the change from a ladder to a half-polygon or viceversa. We provide two examples to illustrate the structure of the resulting configurations in Figure \ref{fig:ladder-polygon}.

\subsection{Graph-theoretical description of multiple sheeted networks}
\label{sec:graph-multiple-sheet}

We now address the question of when a given topological graph (for instance, with topology described in the `ordered \& sewn' representation in section \ref{sec:graph-classification}) admits a description in terms of a multiple-sheeted representation of the kind discussed above. For simplicity we carry out the discussion in terms of compact graphs.

\subsubsection{The general recipe}

For our above examples, it is clear that the main feature is that the set of $E_R$ red edges splits into a number $k$ of $n$-tuples, with $E_R=kn$. All the red edges in a given $n$-tuple have `related' intersections with the green/blue curve, resulting in the ordered ladder structures. This reflects the fact that those edges arise from the same red curve in the $(Z,{\tilde X)}$-plane, but in the $n$ different $X$-sheets. 

Therefore, we only need to quantify in graph-theoretical language the precise meaning of the red edges forming these ladder structures, i.e. the relation between their intersections with the green/blue curve. This can be done as follows. Consider the green/blue path of the graph, by combining green edges with (orientation-flipped) blue edges, and label the vertices (independently of their color) according to their positions along the path with $i=0,\ldots, 2E_R-1$. In general the path contains several multiple components, and their precise ordering is not relevant, as long as the ordering is preserved within each connected component. Note that due to the periodicity of $2E_R$, this labeling is defined up a to cyclic permutation.

As explained in section \ref{sec:graph-rules}, the set of red edges forms a perfect matching, and defines a map from the set of vertices to itself (in particular, off diagonal, sending black nodes to white ones, and viceversa). In our present case, the $n$-sheet condition implies that the ordered set of vertices splits into $2k$ ordered $n$-tuples, such that the perfect matching preserves this splitting. More explicitly, we write the label $i=0,\ldots, 2E_R-1$ of vertices as $i=qn+r$, with $q=0,\ldots, 2k-1$ and $r=0,\ldots, n-1$. Then the perfect matching is a $2E_R\times 2E_R$ matrix $P$, given by the tensor product of a matrix $Q$ acting on the $n\times n$ blocks and a matrix $Q'$ acting inside these blocks, namely
 \beqa
 P_{q_1,r_1;q_2,r_2}=Q_{q_1q_2}Q'_{r_1r_2}
 \eeqa
The matrix $Q$ has a simple interpretation. If we declare the $n$-tuples to define equivalence classes of vertices (assigning them a color by a rule of majority), then $Q$ defined a perfect matching of the resulting set of equivalence classes of vertices. The structure of $Q'$ is also very simple, with value $1$ for entries with $r_1+r_2=n-1$, and zero otherwise \footnote{Note that our convention here is that the perfect matching describes the adjacency of vertices without accounting for orientation, i.e. the entry has equal value for a map from a black to a white vertex or viceversa). It would be easy to introduce signs to include this information, but we will not need to do so.}

For non-compact graphs, in which all red edges are however finite, the same rule applies. For graphs with semi-infinite red edges, the rule only applies to finite edges, and semi-infinite ones are taken to satisfy it automatically.

\begin{figure}[htb]
\begin{center}
\includegraphics[scale=.45]{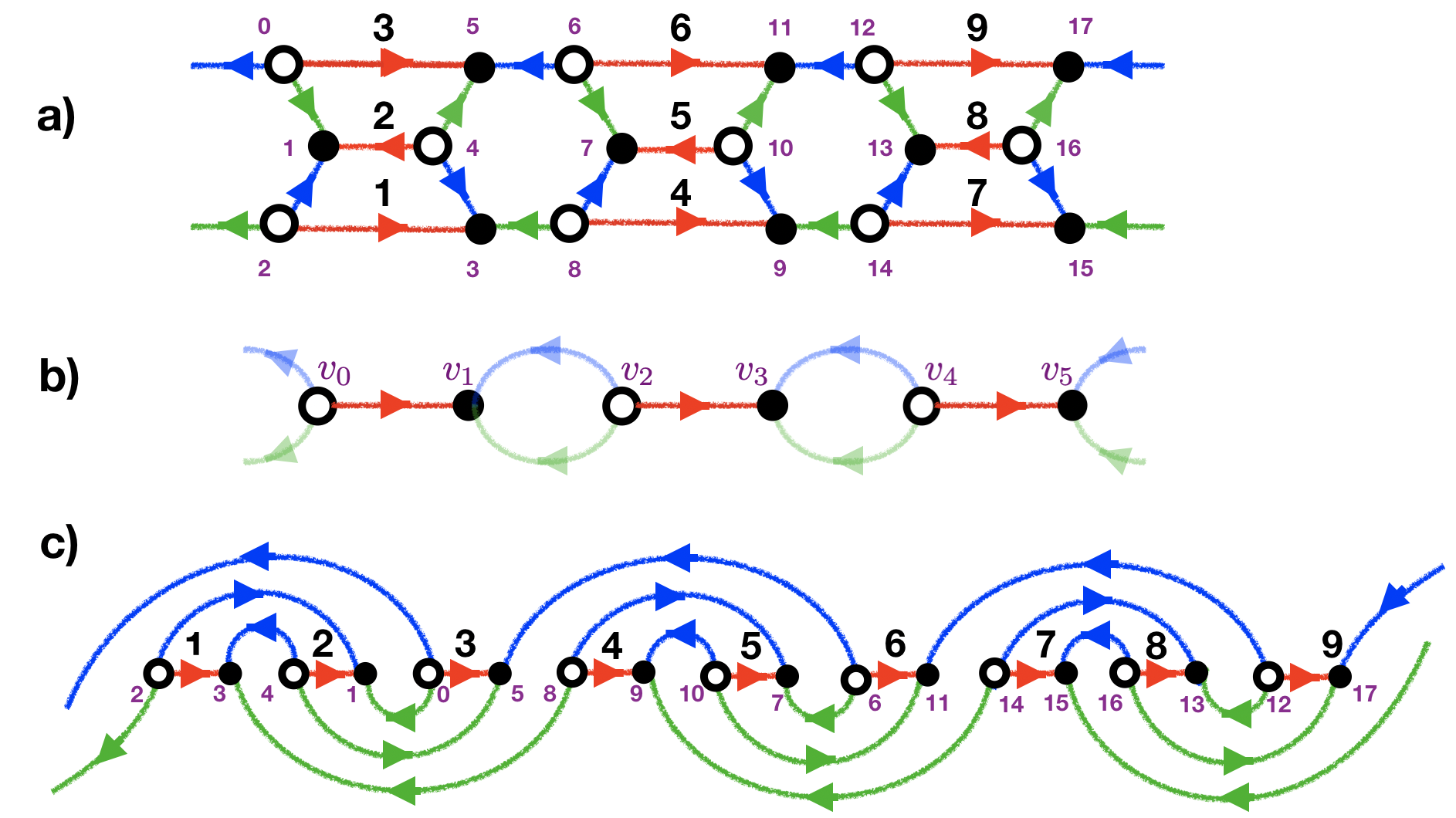}
\caption{\small a) Graph for the non-compact network combining ladders and polygons in Figure \ref{fig:ladder-polygon}a, with vertices labeled (with violet numbers) along the green/blue path, and edges labeled (with black numbers). The graph organizes into 3-tuples of red edges and vertices. b) A graphs for equivalence classes of 3-tuples, as explained in the text. c) Representation in the `ordered \& sewn' form of the general algorithm.}
\label{fig:labeling-ladder1}
\end{center}
\end{figure}

\subsubsection{A concrete example}

The above seemingly complicated recipe is actually easy to check in concrete examples. Let us consider the graph in Figure \ref{fig:ladder-polygon}a, and consider the labeling (in violet) of vertices along its green/blue path as in Figure \ref{fig:labeling-ladder1}a, where we have also labeled the red edges (in black). 

There are $E_R=9$ red edges, which split into 3 different 3-tuples (namely $k=3$, $n=3$). These 3-tuples correspond to the edges $\{1,2,3\}$, $\{4,5,6\}$ and $\{7,8,9\}$. The vertices split into 6 different 3-tuples, corresponding to $\{0,1,2\}$, $\{3,4,5\}$, $\{6,7,8\}$, $\{9,10,11\}$, $\{12,13,14\}$, $\{15,16,17\}$. The relation between the 3-tuples of edges and vertices is already manifest, but we keep on following the general recipe above.

Let us denote $v_q=\{3q,3q+1,3q+2\}$ the equivalence class of vertices in the $q^{th}$ 3-tuple, for $q=0,\ldots, 5$. Then the structure of the matrix $Q$ is encoded in the graph shown in Figure \ref{fig:labeling-ladder1}b (we have depicted the equivalence classes of red edges, but also suggested with faded lines similar equivalence classes of green and blue edges, for later convenience). More explicitly, the matrix $Q$ in the basis $\{v_0,v_2,v_4;v_1,v_3,v_5\}$ has the form
\beqa
Q=\begin{pmatrix} 0 & {\bf 1}_3 \\  {\bf 1}_3 & 0 \end{pmatrix}
\eeqa
Then, within each equivalence class, the structure of the matrix $Q'$ in the basis of indices $r\in\{0,1,2\}$ is 
\beqa
Q'=\begin{pmatrix} 0 & 0 & 1 \\  0 & 1 & 0 \\ 1 & 0 & 0 \end{pmatrix}
\eeqa
namely, the only non-zero entries are those with indices adding up to $n-1=2$.

Let us also mention that in this procedure, it is clear that the auxiliary graph of equivalence classes in Figure \ref{fig:labeling-ladder1}b is precisely that encoding the structure of the superpotential proportional to $\Lambda$, i.e. the graph that would result if the configuration were single-sheeted.

Finally, for illustration, Figure \ref{fig:labeling-ladder1}c we also provide the representation of the graph in the `ordered \& sewn' form of the general algorithm. Notice that, although the latter figure seems to display some nice symmetry properties, there is no obvious pattern reflecting the possibility to describe it as an $n$-sheeted diagram, except for the structure described above. 

We refrain from discussing further examples at this point, and move on to a brief discussion of the possibility of ${\tilde X}$-sheets.

\subsection{The ${\tilde X}$-sheets}
\label{sec:xtilde-sheets}

We now consider the effect of adding extra ${\tilde X}$-sheets, in a way similar to the $X$-sheets in the previous section. Hence, we consider several sheets for $X$ and ${\tilde X}$ by promoting (\ref{fn-x-sheets}) to
\beqa
f_n(X)g_{2m+1}(\tilde X)=0\, .
\eeqa
Note that because the superpotential term which goes with ${\tilde \Lambda}$ must be an odd function of ${\tilde X}$, we must have and odd number of ${\tilde X}$-sheets (and they must be arranged in a $\IZ_2$ symmetric way), namely $g_{2m+1}(\tilde X)$ is an odd function. We denote by $X_i$, $i=1,\ldots,n$ the $n$ roots of $f_n(X)$, and by ${\tilde X}_j$, $j=-m,\ldots , 0,\ldots m$ the $2m+1$ roots of $g_{2m+1}({\tilde X})$, with ${\tilde X}_{-j}=-{\tilde X}_j$ and ${\tilde X}_0=0$.

Let us focus on a concrete illustrative example. For instance, we supplement the above equation with
\beqa
{\tilde X}^2-X^2-Z=0\, .
\eeqa
We start by focusing on the branches with ${\tilde X}= {\tilde X}_j$. On these branches, the curve on the $(X,Z)$-plane is
\beqa
Z={\tilde X}_j^2-X^2\, .
\eeqa
These are $(2m+1)$ green/blue parabolas (at ${\tilde X}={\tilde X}_j$), open towards $Z\to -\infty$. Note that in the quotient by ${\tilde X}\to -{\tilde X}$ we get only $(m+1)$ independent curves.

The branches with $X=X_i$ produce curves in the $({\tilde X},Z)$-plane given by
\beqa
Z= {\tilde X}^2-X_i^2\, .
\eeqa
These are $n$ red parabolas (at $X=X_i$),  open towards $Z\to +\infty$.

The two sets of parabolas intersect at $X=X_i$, ${\tilde X}={\tilde X}_j$, $Z=  {\tilde X}_j^2-X_i^2$. In the covering space we have $n(2m+1)$ intersection points, $n$ of them at ${\tilde X}=0$ and $2nm$ at $\IZ_2$ symmetric points away from ${\tilde X}=0$. Hence, after the $\IZ_2$ quotient we get $n+nm=n(m+1)$ points. Namely each of the $n$ parabolas in the $X$-sheets intersect each of the $(m+1)$ parabolas in the ${\tilde X}$-sheets.

We must note that the each curves on the ${\tilde X}$-sheets at ${\tilde X}\neq 0$ is not mapped to itself under the $\IZ_2$, rather they are exchanged in pairs. This means that they do not correspond to blue of green curves (which are mapped to themselves, so they must sit at ${\tilde X}$), rather they correspond to red lines.

\begin{figure}[htb]
\begin{center}
\includegraphics[scale=.5]{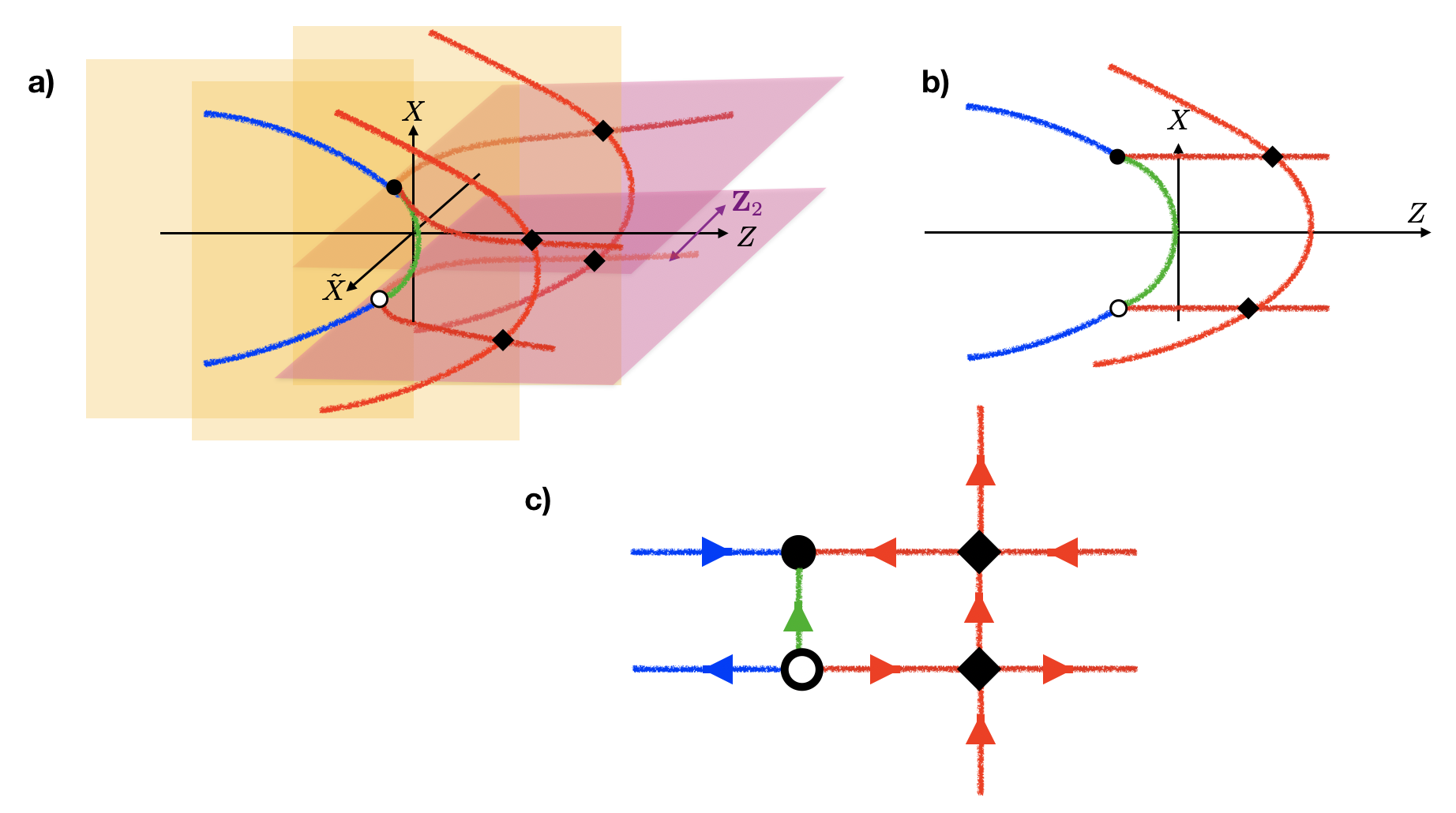}
\caption{\small Graph for a configuration with multiple $X$- and ${\tilde X}$-sheets, for the example of $n=2$, $m=1$. Figure a) shows the configuration in the covering space, b) a depiction of the configuration in the quotient space, c) a graph representation. For clarity we have indicated trivalent junctions with (black or white) dots, and intersections with 4 branches with black diamonds.}
\label{fig:double-sheets}
\end{center}
\end{figure}

On a related note, the intersections of the parabolas which are away from ${\tilde X}=0$ do not correspond to trivalent junctions, but to junctions with trivial chiral flow: the two red lines do not end on the other parabola, but intersect it and continue. We just get two red theories that `pass each other' without exchanging chiral matter. As already explained, we represent such non-chiral junctions with a diamond. In Figure \ref{fig:double-sheets} we provide one example of the above kind of configurations, with 2 $X$-sheets and 3 ${\tilde X}$-sheets.

The above remarks hold for any generic model with multiple  ${\tilde X}$-sheets. Hence, the introduction of extra ${\tilde X}$-sheets leads to the presence of additional branches at ${\tilde X}\neq 0$, whose junctions are necessarily non-chiral. 
Since we are not interested in graphs with non-chiral junctions, we will not consider the possibility of extra ${\tilde X}$-sheets any further.

\section{Further applications}
\label{sec:applications}

In this section we consider several interesting extensions of the ideas presented above, related to extending the concept of network to higher-dimensional configurations, or to regard those constructed above in a new light, as possibly leading to new compactifications to lower spacetime dimensions.

\subsection{Higher dimensional networks}
\label{sec:higher}

We have so far focused on one-dimensional networks and their characterization by graphs. In this section we emphasize that one can introduce dependence of the $(0,1)$ worldsheet superpotential in more spacetime dimensions and get  interesting higher-dimensional configurations. These describe interesting processes, for instance, as one moves in one direction, the network can nucleate conjugate junction pairs, which are subsequently annihilated. Such phenomena can be useful to describe on-shell solutions in which the dynamics of the system can be addressed. We leave a more systematic discussion of higher-dimensional configurations for future work, and simply provide some interesting examples.

\subsubsection{The dynamical transition}
\label{sec:dynamical-transition}

\begin{figure}[htb]
\begin{center}
\includegraphics[scale=.35]{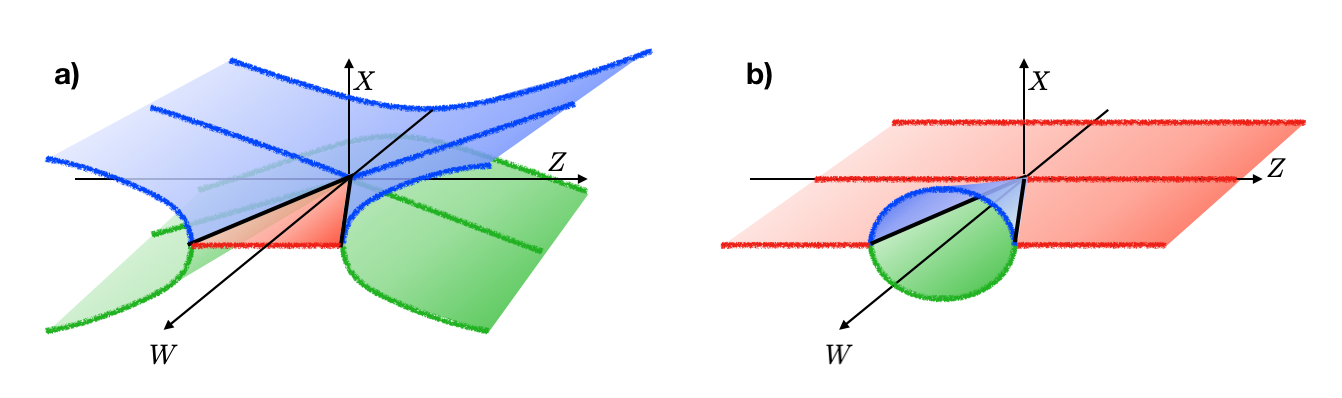}
\caption{\small A two-dimensional network can be regarded as a  dynamical version of the transitions in  section \ref{sec:transition}. As in previous cases, the red and blue/green regions actually have a more symmetric look if we would depict the direction ${\tilde X}$ (which we are projecting in the figure) and worked on the covering space, before the orbifold identification of the two red branches.}
\label{fig:nucleation}
\end{center}
\end{figure}

We can now consider a dynamical realization of the transition in section \ref{sec:transition}. The idea is simple: just promote the parameter $a$ to a function of some 10d spacetime coordinate $W$, different from $Z$. Although a real dynamical realization would require to study time-dependent configurations, we will restrict to spatial dependences (yet still refer to the configuration as `dynamical'). We therefore can consider
\beqa
W=\Lambda({\tilde X}^2-X^2-f(Z,W))+ {\tilde \Lambda} X{\tilde X}\, ,
\label{codim2-general-superpotential}
\eeqa
with
\beqa
f(Z,W)=\pm (Z^2-W)\,  .
\label{parabola3}
\eeqa
Then, for $W<0$ we recover the behaviour of $a<0$ in section \ref{sec:transition} (trivial networks, because there are no zeroes for any value of $Z$), while for $W>0$ we obtain the behaviour of $a>0$ (non-trivial network, with two zeroes at two values of $Z$ with opposite slope of $\partial_Z f$). The configuration can be regarded as the dynamical nucleation of an intermediate branch, as follows:

In the case of overall negative sign in (\ref{parabola3}), we start (at $W\ll 0$) with two disjoint 10d theories (corresponding to the $Spin(32)/\IZ_2$ and $SO(16)\times SO(16)$ theories), and at some point ($W=0$) a bubble of a new 10d theory  (an $E_8\times E_8$ theory) appears, and grows indefinitely (at $W\gg 0$). This is shown in Figure \ref{fig:nucleation}a.

In the case of overall positive sign in (\ref{parabola3}), we start (at $W\ll 0$) with one 10d theory (an $E_8\times E_8$ theory), and at some point ($W=0$) a bubble of two new disconnected 10d theories  appears (corresponding to the $Spin(32)/\IZ_2$ and $SO(16)\times SO(16)$ theories), and grows indefinitely (at $W\gg 0$). This is shown in Figure \ref{fig:nucleation}b.

Notice that, if we focus on the $Z=0$ slice, the configurations look like the basic junctions in section \ref{sec:conjugate}, with the appearance of the junction associated to the linear term $W$. This motivates the construction in the next section.

\subsubsection{Bubbling}

\begin{figure}[htb]
\begin{center}
\includegraphics[scale=.30]{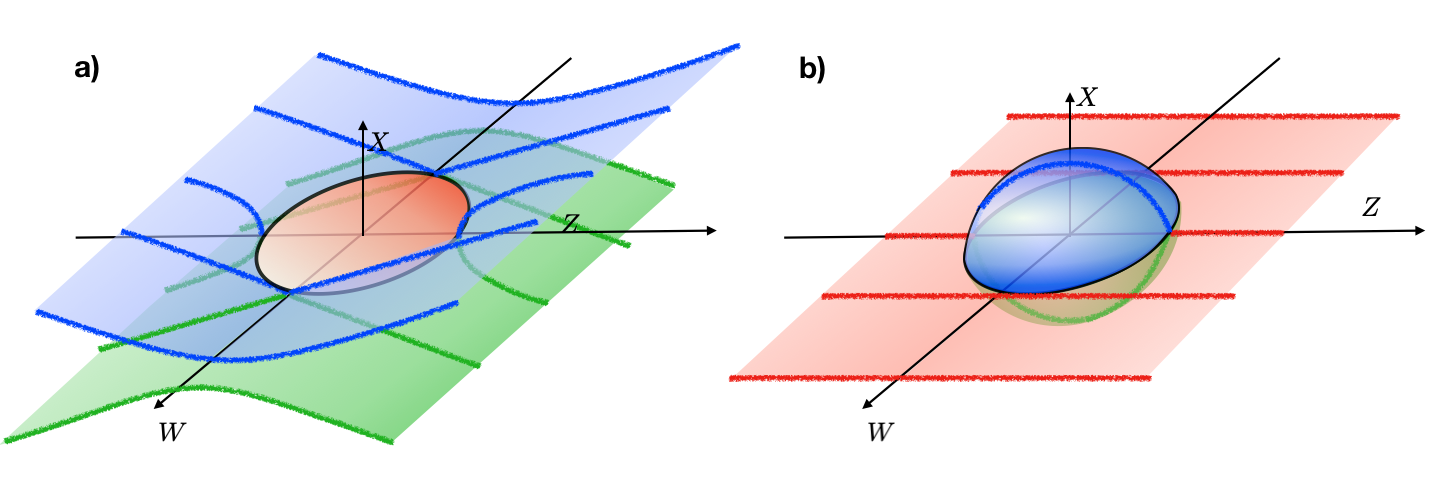}
\caption{\small (a) Bubble of a 10d $E_8\times E_8$ heterotic theory inside a configuration of two disjoint 10d $Spin(32)/\IZ_2$ and $SO(16)\times SO(16)$ theories. (b) The reverse configuration, a bubble of two disjoint 10d $Spin(32)/\IZ_2$ and $SO(16)\times SO(16)$ theories in a 10d $E_8\times E_8$ heterotic theory.}
\label{fig:codim2-bubbles}
\end{center}
\end{figure}

\begin{figure}[htb]
\begin{center}
\includegraphics[scale=.35]{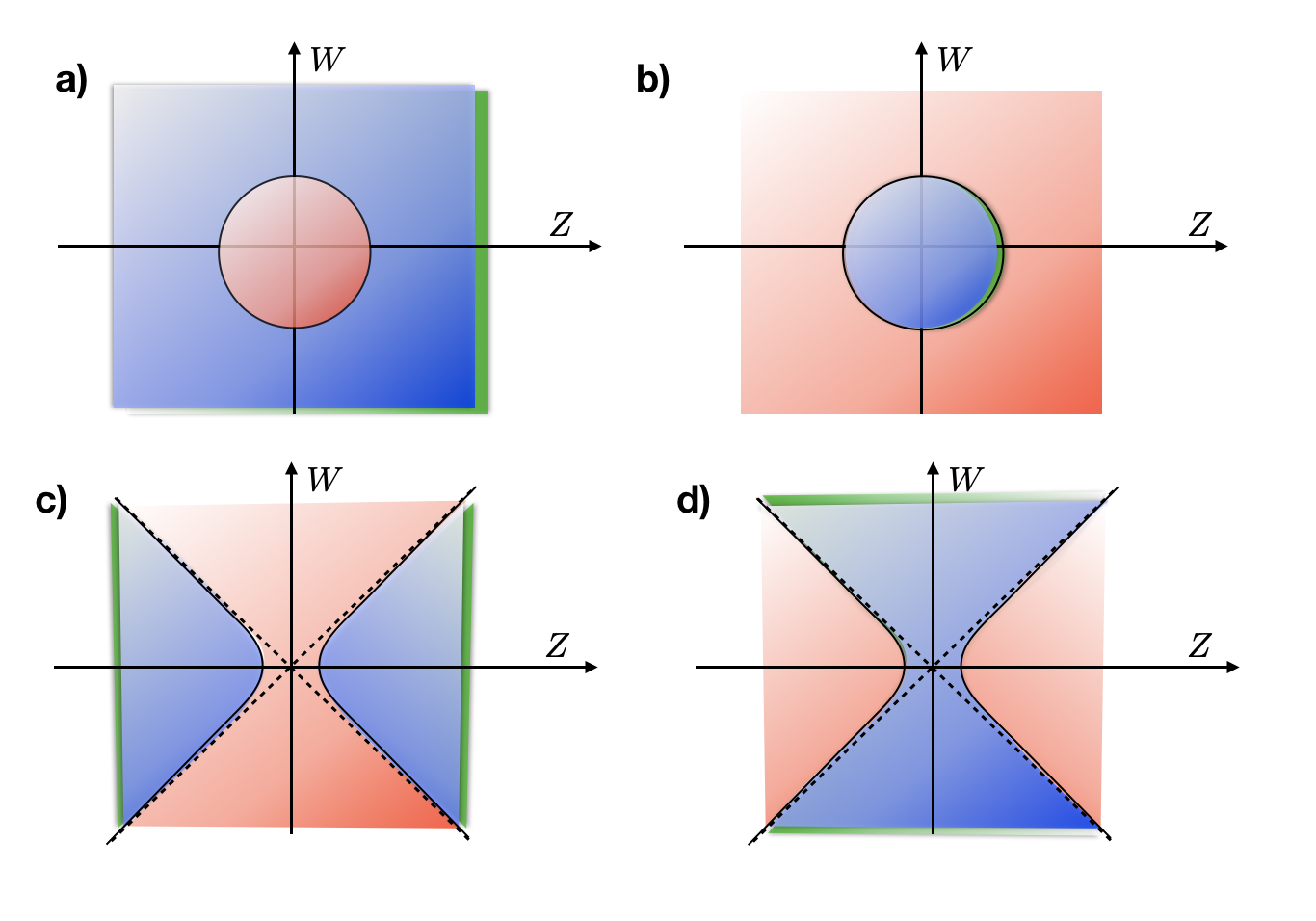}
\caption{\small (a) and (b) shown the bubbles in Figure \ref{fig:codim2-bubbles}a,b in the $(Z,W)$ plane. Figures (c), (d) show the configuration for opposite sign of $Z^2$ and $W^2$, for one choice of sign of $a$. For the opposite choice of sign of $a$, we get similar picture with the exchange of roles of $Z$ and $W$. }
\label{fig:codim2-bubbles-hyperbola}
\end{center}
\end{figure}

Amusingly, we can consider the above dependence on $W$ in (\ref{parabola3})  to correspond to the linear approximation to a more general function. For instance, we can promote it to a parabola, namely $W\to \pm W^2$ (plus an additive constant $a$). We have two main cases, depending on whether $Z^2$ and $W^2$ have equal or opposite sign. If we have
\beqa
f(Z,W)=\pm (Z^2+W^2-a)\, .
\label{parabola4}
\eeqa
Then, for $a>0$ the zeros are located at the circle 
\beqa
Z^2+W^2=a\, .
\eeqa
Clearly, as we move in $W$, the configuration corresponds to starting in a configuration at $W\ll -\sqrt{a}$, then at $W=-\sqrt{a}$ we nucleate a bubble, i.e. a region of the other configuration, and recollapse it at $W=+\sqrt{a}$, after which we continue with the original configuration at $W\gg \sqrt{a}$. In other words, we have a network of the same kind in both planes $W=0$ and $Z=0$. The configuration describes a bubble of one universe inside an ambient universe. The overall change of sign determines the nature of the ambient universe. This is depicted in Figure \ref{fig:codim2-bubbles}a (for negative overall sign) and (b) (for positive overall sign), see also Figure \ref{fig:codim2-bubbles-hyperbola}a,b.

If we take opposite signs for $Z^2$ and $W^2$ we have
\beqa
f(Z,W)=\pm (Z^2-W^2-a)\, .
\label{parabola5}
\eeqa
Then, independently of the sign of $a$ we have non-trivial zeroes, along the hyperbola
\beqa
Z^2-W^2=a\, .
\eeqa
The configurations look like one 10d universe of one kind extending infinitely in one direction ($Z$ or $W$) and bounded in the other by two 10d universes of the other kind, which extend infinitely in the same direction ($Z$ or $W$) and semi-infinitely in the other direction ($W$ or $Z$). Topologically this is just like one of the networks in Figure \ref{fig:network}. This is shown in figure \ref{fig:codim2-bubbles-hyperbola}, in a more schematic depiction.

\subsubsection{Compact bubbles}
\label{sec:compact-bubbles}

\begin{figure}[htb]
\begin{center}
\includegraphics[scale=.35]{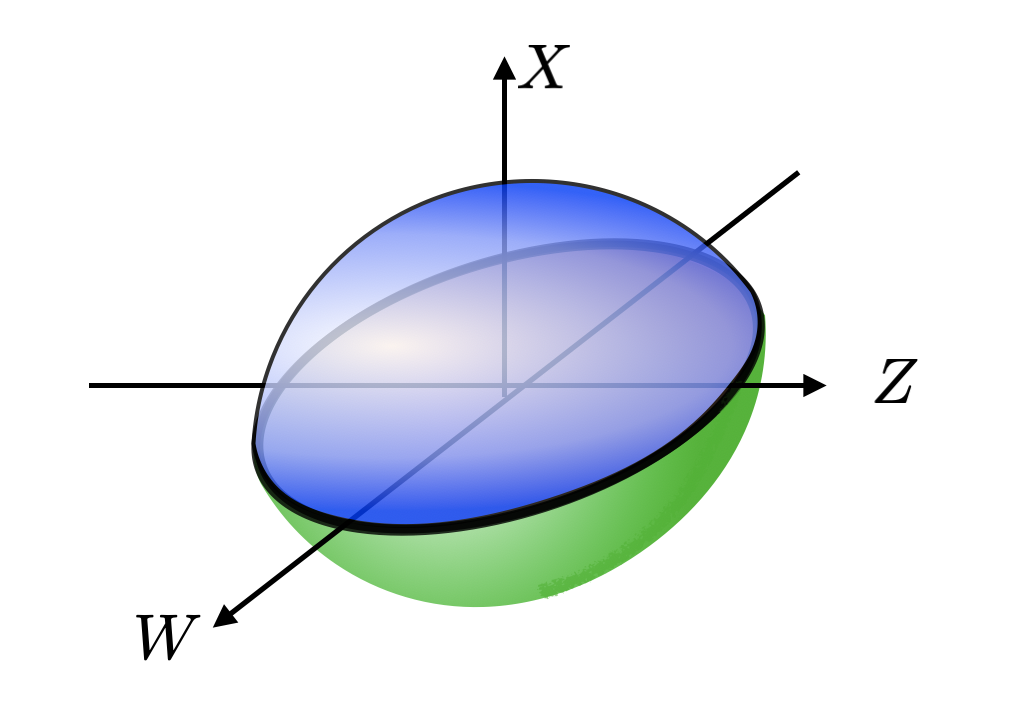}
\caption{\small The 2d ball configuration corresponds to the junction spanning an $\IS^1$ which bounds three individual disks, for the three different colors. The blue and green ones, shown in the figure, form an $\IS^2$ in the $(Z,X,W)$-space, while the one is described as an $\IS^2$ in the $\IZ_2$ covering $(Z,{\tilde X},W)$-space. The three share the boundary, given by an $\IS^1$ in the $(Z,W)$-plane.}
\label{fig:ball}
\end{center}
\end{figure}

Let us now show a simple example of 2-dimensional compact  network. In order to achieve compactness, we simply modify (\ref{codim2-general-superpotential}) by flipping the sign of the $X^2$ term and take
\beqa
W=\Lambda({\tilde X}^2+X^2+Z^2+W^2-a)+ {\tilde \Lambda} X{\tilde X}\; ,{\rm for}\; a>0\, .
\label{supo-ball}
\eeqa
For ${\tilde X}=0$ we obtain a green/blue 2d area described by an $\IS^2$
\beqa
X^2+Z^2+W^2=a\quad ,\quad {\tilde X}=0\, .
\label{green-blue-sphere}
\eeqa
However, this is split in two hemispheres by the red 2d area defined by $X=0$, which in the $\IZ_2$ covering space corresponds to the $\IS^2$
\beqa
{\tilde X}^2+Z^2+W^2=a\quad ,\quad {\tilde X}=0
\eeqa
After the $\IZ_2$ quotient, we obtain a disk, whose boundary $\IS^1$ splits the $\IS^2$ (\ref{green-blue-sphere}) into a green hemisphere and a blue hemisphere. The configuration, which we dub the 2d ball,  is depicted in Figure \ref{fig:ball}. It is the 2-dimensional version of Figure \ref{fig:compact-network}, and can be regarded as describing a peculiar compactification to 8d, as we further discuss in section \ref{sec:compactifications}.

\subsubsection{Generalizations}

We can easily generalize the above story to any arbitrary conic of any dimension. We just choose some 10d spacetime coordinates $Z_1,\ldots, Z_n$ and choose $f(Z_1,\ldots , Z_n)$ a quadratic expression to introduce in the superpotential. 

In general we can take an arbitrary function and look at the level set $f(Z_1,\ldots , Z_n)$. This will define a set of codimension 1 walls, some of which extend infinitely, others are compact. Upon crossing a wall one jumps from a red region to a green/blue region. We refrain from discussing explicitly these more involved configurations, leaving them as an exercise for the interested reader.

\subsection{Compact networks as singular compactifications}
\label{sec:compactifications}

In this section we reinterpret the configurations described by compact networks as singular compactifications. For concreteness we focus on the one-dimensional networks associated to compact graphs, which are interpreted as compactifications to 9d (generalization to higher-dimensional networks is straightforward). The remarkable new feature is that the kind of 10d heterotic theory in the higher-dimensional viewpoint changes as one moves in the compact dimension. In addition, the non-trivial chiral flow implies that different chiral fields, which propagate on different subgraphs of the complete graph, are effectively compactified on different internal spaces.

Let us consider for concreteness the compact bubble diagram in Figure \ref{fig:compact-network}. It describes a configuration with 9d non-compact dimensions, and one extra dimension which spans a singular space described by a graph. The latter is defined by three 10d heterotic theories, each effectively compactified on its own segment, and with suitable junction conditions at the endpoints to glue them. 

We will not discuss in details the junction conditions, but it is expected that they will lead to e.g. a single zero mode  for the graviton (morally the constant mode in the internal singular space, possibly distorted by the effective tension of the junction), leading to a 9d gravitational theory with a finite Planck scale. We may also expect that other fields which exist in all three theories, like the dilaton, 2-form and $SO(16)\times SO(16)$ gauge bosons, will also survive in the resulting 9d theory. We emphasize that all those fields propagate in the singular space defined by the full compact network.

There are other fields which are only defined in one of the theories. For instance, the gauge bosons in the spinor representations of $SO(16)\times SO(16)$ exist only in the 10d $E_8\times E_8$ heterotic branch, and the gauge bosons in the $({\bf 16},{\bf 16})$ of $SO(16)\times SO(16)$ exist only in the 10d  $Spin(32)/\IZ_2$ heterotic branch. Therefore these fields are compactified in the corresponding segment, with boundary conditions effectively determined by the junction conditions. We expect that these will remove the zero mode of the corresponding fields, so that there are no 9d massless states associated to them.

Finally, the 10d chiral fields of the different theories propagate across the junction according to the chiral flow, so that they live in a compact space given by sub-graph of the complete graph. In particular, they correspond to the two-colored paths $G{\ov B}$, $B{\ov R}$ and $R{\ov G}$ (or their orientation flipped versions) introduced in section \ref{sec:graph-rules}. Each of these fields is defined on a circle in the present example, albeit different ones for different fields. One may then expect that they will lead to massless modes upon compactification, by taking the constant zero mode in the internal space. However, since these fields are fermions, whether they provide massless modes or not depends on the choice of spin structure in the corresponding circle. The minimal assumption is that they will have odd spin structure and that they do not produce additional massless 9d fields. In particular this avoids the potentially problematic situation of finding a 9d theory with massless gravitinos but no supersymmetry, which is in general not expected to exist.

The fact that different fields propagate in different geometries is very reminiscent of the DRP and CRP boundary conditions in M-theory on $\IS^1\vee\IS^1$ \cite{Baykara:2026gem}\cite{Altavista:2026evd}\cite{Baykara:2026vdc,Altavista:2026brr,Dasgupta:2026maq,Basile:2026trt,Kamal:2026msr} A difference is that in our network setup this is a classical (albeit singular) geometry, while in the $\IS^1\vee\IS^1$ setup it is meant to represent a quantum geometry. However,  there are recent attempts \cite{Basile:2026trt} to derive some properties of the $\IS^1\vee \IS^1$ compactification form its geometrization  as a junction. We leave the exploration of the relation of network compactifications with this kind of quantum geometries as an open question for future research.

We finish with a final observation. The realization of compact higher-dimensional networks, c.f. section \ref{sec:compact-bubbles}, provides a generalization of the above idea, which may allow to introduce further ingredients. For instance, by considering the network to support heterotic theories defined on $2n$-dimensional spaces bounded by junctions on their $(2n-1)$-dimensional boundary, it may be possible to turn on gauge bundles with non-trivial $\tr  F^n$ on some of the theories, in such a way that they  help to stabilize the configuration. This is in analogy with the introduction of fluxes stabilizing otherwise unstable sphere compactifications, as mentioned in section \ref{sec:stability}. We leave this interesting direction for future work.

\section{Conclusions}
\label{sec:conclusions}

In this work we have initiated the analysis of networks built out of junctions of 10d string theories, considering the case study of those arising from the trivalent junction of the three 10d non-tachyonic heterotic theories. The configurations as such are admittedly not expected to be stable, and complicated networks are likely decaying into simpler ones. Nevertheless, it is reasonable to expect that they may be dressed with extra ingredients (like spacetime dependence, or stabilization mechanisms such as branes or fluxes) providing their realization as full solutions of the equations of motion. Hence, we expect that the techniques and results we have obtained provide a good view of the possible configuration space in which to seek the implementation of these extra ingredients.

We have characterized the topologically allowed networks via elementary graph theory, provided a realization of the graph embedding as an algebraic variety, and shown that there is a natural generalization of the construction in \cite{Tachikawa:2026top} allowing to realize the network in terms of a 2d $(0,1)$ heterotic worldsheet theory. Although we have focused on junctions of 10d non-tachyonic heterotic theories, they also hold for the more general junctions relating  three worldsheet CFTs $T$, $T/\IZ_2$ and $(T\times q)/\IZ_2$ \cite{Tachikawa:2026top}, c.f. footnote \ref{foot:general-junction}.

We have provided many explicit examples illustrating these constructions. We have moreover shown that the constructions admit simple generalizations to higher-dimensional networks, which may lead to their application in physical contexts, such as cosmological bubble nucleation.

Finally, we have made the interesting observation that compact networks can be regarded as a novel kind of compactification, in with the underlying 10d theory changes as one moves in the compact space. Very intriguingly, a built-in feature of this setup is that different fields can propagate on different sub-graph geometries of the full network. This is tantalizingly reminiscent of similar behavior in the context of M-theory on $\IS^1\vee \IS^1$ (and quotients and duals thereof) \cite{Baykara:2026gem}\cite{Altavista:2026evd}\cite{Baykara:2026vdc,Altavista:2026brr,Dasgupta:2026maq,Basile:2026trt,Kamal:2026msr}, so it may provide a handle on the otherwise ad hoc choices in these latter quantum compactifications.

There are many open directions to pursue in future work. Some of them are:

\begin{itemize}

\item A most prominent question is the dynamics of the networks. This can be approached from the perspective of the RG flow on the worldsheet and its reflection in e.g. time-dependent backgrounds describing on-shell solutions. Or, in a complementary way, from the spacetime point of view in interplay with the introduction of possible stabilization mechanisms. This is in close analogy with other non-supersymmetric systems, such as brane-antibrane configurations, which were initially considered merely as off-shell configurations displaying new physics, but subsequently turned not only into established model building ingredients, but also into the cornerstone of our present K-theoretical understanding of D-branes and RR fields.

\item We have focused on networks built from the trivalent heterotic junction just as a case study of a much more general problem of understanding networks of  quantum gravity theories. They provide a possibly well controlled arena, in particular because the chiral flow ensures there is no need of strong coupling to gap any chiral field content.  It would be interesting to extend our analysis to networks of other theories, for instance networks built from the 4-valent bouquet in \cite{Altavista:2026edv} of the 10d type IIB, type I, $USp(32)$, and $U(32)$ 0'B theories. The structure of the graphs is a generalization of what we have studied:  bipartite graphs with 4-colored edges, and vertices involving one edge of each color. Several concepts like the two-colored paths generalize directly, and describe the chiral flow in the network. Finding a worldsheet description of the resulting networks would provide non-trivial tools to understand the behavior of orientifold planes and D-branes across the network.

\item The intriguing resemblance of some features of compact networks with the proposal of new quantum geometries like $\IS^1\vee \IS^1$ deserves further study. In particular, to determine if the classical configurations of compact networks are related to configurations with sizes frozen at small scales (in analogy with e.g. Gepner or LG models correspond to CY compactifications with stringy size), or alternatively, if there exist quantum geometries which admit a large volume limit in which they become networks of the corresponding theory.

\item More speculatively, one may wonder about a more fundamental role of networks in the structure of spacetime. In particular one may consider configurations described by very dense graphs, with edges of very small length, and seek a long-wavelength description of the corresponding dynamics, as providing a sort of continuum limit. We envision an interesting interplay of these systems with quantum information theory, the geometrization of entanglement via graphs, and the grand questions about the emergence of spacetime. 

\end{itemize}

We hope to report on progress on these and other directions in the future.

\section*{Acknowledgments}

We are pleased to thank Ivano Basile, Ralph Blumenhagen, Markus Dierigl, Dieter L\"ust, Carmine Montella, Fernando Quevedo, and Ignacio Ruiz for useful discussions. A. U. thanks the Max Planck Institute and the Ludwig-Maximilans University in Munich, for providing a friendly and stimulating atmosphere in Ralph Blumenhagen's Fest, where part of these results were communicated. This work is supported through the grants CEX2020-001007-S, PID2021-123017NB-I00 and  ATR2023-145703 funded by MCIN/AEI/10.13039/501100011033 and by ERDF A way of making Europe. 
C. A. is supported by the fellowship LCF/BQ/DFI25/13000111 from ``La Caixa'' Foundation (ID 100010434). E. A. is supported by the fellowship LCF/BQ/DI24/12070005 from ``La Caixa'' Foundation (ID 100010434). R. A. acknowledges support from the ERC Starting Grant QGuide- 101042568- StG 2021, the Deutsche Forschungsgemeinschaft through the Collaborative Research Center 1624 “Higher Structures, Moduli Spaces and Integrability” and the Deutsche Forschungsgemeinschaft under Germany’s Excellence Strategy EXC 2121 Quantum Universe 390833306. C. W. is supported by program PIPF-2024/TEC-34293 from Comunidad de Madrid.

\bibliographystyle{JHEP}
\bibliography{refs}


\newpage


\end{document}